\documentclass[12pt,showpacs,preprintnumbers,amsmath,amssymb]{revtex4}
\usepackage[english]{babel}
\usepackage[T1]{fontenc}
\usepackage{bbm}
\usepackage{epsfig}
\usepackage{psfrag}
\usepackage{color}

	\newcommand{\iit}{\it}
	\newcommand{\bbf}{\bf}

	\newcommand{\be}{\begin{equation}}
	\newcommand{\ee}{\end{equation}}

	\newcommand{\PHI}{{\Phi}}
	\newcommand{\EXP}[1]{\mathrm{e}^{#1}}

	\newcommand{\dmat}{\mathrm{d}}
	
	\newcommand{\diff}{\frac{1}{\rho}\partial_\rho\left(\rho\partial_\rho\right)}
	
	\newcommand{\D}{{\cal D}}

	\newcommand{\thetaw}{\theta_{\mbox{\tiny W}}}

	\newcommand{\vano}{v_{\mbox{\tiny ANO}}}
	\newcommand{\fano}{f_{\mbox{\tiny ANO}}}

	\newcommand{\okm}{{\omega,\kappa,m}}

	\newcommand{\phase}{\Xi}
	\newcommand{\imatt}{i}	

	\newcommand{\Zz}{\mathcal{Z}_0}
	\newcommand{\Zp}{\mathcal{Z}_+}
	\newcommand{\Zm}{\mathcal{Z}_-}
	\newcommand{\Ze}{\mathcal{Z}_\eta}
	\newcommand{\Az}{\mathcal{A}_0}
	\newcommand{\Ap}{\mathcal{A}_+}
	\newcommand{\Am}{\mathcal{A}_-}
	\newcommand{\Ae}{\mathcal{A}_\eta}
	
	\newcommand{\Wpz}{\mathcal{W}_0^+}
	\newcommand{\Wpp}{\mathcal{W}_+^+}
	\newcommand{\Wpm}{\mathcal{W}_-^+}
	\newcommand{\Wmz}{\mathcal{W}_0^-}
	
	\newcommand{\Wmp}{\mathcal{W}_+^-}
	\newcommand{\Wmm}{\mathcal{W}_-^-}
	\newcommand{\Hop}{h_1^+}
	\newcommand{\Hom}{h_1^-}
	\newcommand{\Htp}{h_2^+}
	\newcommand{\Htm}{h_2^-}
	\newcommand{\Zpm}{\mathcal{Z}_\pm}
	\newcommand{\Zmp}{\mathcal{Z}_\mp}
	\newcommand{\Apm}{\mathcal{A}_\pm}
	\newcommand{\Amp}{\mathcal{A}_\mp}
	\newcommand{\Wpmp}{\mathcal{W}_+^\pm}
	\newcommand{\Wpmm}{\mathcal{W}_-^\pm}
	\newcommand{\Wpmpm}{\mathcal{W}_\pm^\pm}
	\newcommand{\Wmpmp}{\mathcal{W}_\mp^\mp}
	\newcommand{\Wpmz}{\mathcal{W}_0^\pm}
	\newcommand{\Wmpz}{\mathcal{W}_0^\mp}
	\newcommand{\Wpe}{\mathcal{W}_\eta^+}
	\newcommand{\Wme}{\mathcal{W}_\eta^-}
	\newcommand{\Wpme}{\mathcal{W}_\eta^\pm}
	
	\newcommand{\Hopm}{h_1^\pm}
	\newcommand{\Htpm}{h_2^\pm}

	\newcommand{\Gza}{\Gamma_{\mbox{\tiny $\mathcal{Z}\mathcal{A}$}}}
	\newcommand{\Gzwp}{\Gamma_{\mbox{\tiny $\mathcal{Z}\mathcal{W}^+$}}}
	\newcommand{\Gzwm}{\Gamma_{\mbox{\tiny $\mathcal{Z}\mathcal{W}^-$}}}
	\newcommand{\Gzh}{\Gamma_{\mbox{\tiny $\mathcal{Z}\mathcal{H}$}}}
	\newcommand{\Gawp}{\Gamma_{\mbox{\tiny $\mathcal{A}\mathcal{W}^+$}}}
	\newcommand{\Gawm}{\Gamma_{\mbox{\tiny $\mathcal{A}\mathcal{W}^-$}}}
	\newcommand{\Gah}{\Gamma_{\mbox{\tiny $\mathcal{A}\mathcal{H}$}}}
	\newcommand{\Gww}{\Gamma_{\mbox{\tiny $\mathcal{W}\mathcal{W}$}}}
	\newcommand{\Gwph}{\Gamma_{\mbox{\tiny $\mathcal{W}^+\mathcal{H}$}}}
	\newcommand{\Gwmh}{\Gamma_{\mbox{\tiny $\mathcal{W}^-\mathcal{H}$}}}
	\newcommand{\Gxy}{\Gamma_{\mbox{\tiny $xy$}}}

	\newcommand{\DZ}{{\Delta}_{\mbox{\tiny $\mathcal{Z}$}}}
	\newcommand{\DA}{{\Delta}_{\mbox{\tiny $\mathcal{A}$}}}
	\newcommand{\DWp}{{\Delta}_{\mbox{\tiny $\mathcal{W}^+$}}}
	\newcommand{\DWm}{{\Delta}_{\mbox{\tiny $\mathcal{W}^-$}}}
	\newcommand{\DWpm}{{\Delta}_{\mbox{\tiny $\mathcal{W}^\pm$}}}
	\newcommand{\DHH}{{\Delta}_{\mbox{\tiny $\mathcal{H}$}}}

	\newcommand{\Dzz}{{\Delta}^{\mbox{\tiny $\mathcal{Z}$}}_{\mbox{\tiny $0$}}}
	\newcommand{\Dzp}{{\Delta}^{\mbox{\tiny $\mathcal{Z}$}}_{\mbox{\tiny $+1$}}}
	\newcommand{\Dzm}{{\Delta}^{\mbox{\tiny $\mathcal{Z}$}}_{\mbox{\tiny $-1$}}}
	\newcommand{\Daz}{{\Delta}^{\mbox{\tiny $\mathcal{A}$}}_{0}}
	\newcommand{\Dap}{{\Delta}^{\mbox{\tiny $\mathcal{A}$}}_{\mbox{\tiny $+1$}}}
	\newcommand{\Dam}{{\Delta}^{\mbox{\tiny $\mathcal{A}$}}_{\mbox{\tiny $-1$}}}
	\newcommand{\Dwpmz}{{\Delta}^{\mbox{\tiny $\mathcal{W}^\pm$}}_{\mbox{\tiny $0$}}}
	\newcommand{\Dwpmp}{{\Delta}^{\mbox{\tiny $\mathcal{W}^\pm$}}_{\mbox{\tiny $+1$}}}
	\newcommand{\Dwpmm}{{\Delta}^{\mbox{\tiny $\mathcal{W}^\pm$}}_{\mbox{\tiny $-1$}}}
	\newcommand{\Dhop}{{\Delta}^{\mbox{\tiny $h_1$}}_{\mbox{\tiny $+$}}}
	\newcommand{\Dhom}{{\Delta}^{\mbox{\tiny $h_1$}}_{\mbox{\tiny $-$}}}
	\newcommand{\Dhtp}{{\Delta}^{\mbox{\tiny $h_2$}}_{\mbox{\tiny $+$}}}
	\newcommand{\Dhtm}{{\Delta}^{\mbox{\tiny $h_2$}}_{\mbox{\tiny $-$}}}

	\newcommand{\Dze}{{\Delta}^{\mbox{\tiny $\mathcal{Z}$}}_{\mbox{\tiny $\eta$}}}
	\newcommand{\Dae}{{\Delta}^{\mbox{\tiny $\mathcal{A}$}}_{\mbox{\tiny $\eta$}}}
	\newcommand{\Dwpme}{{\Delta}^{\mbox{\tiny $\mathcal{W}^\pm$}}_{\mbox{\tiny $\eta$}}}
	\newcommand{\Dhopm}{{\Delta}^{\mbox{\tiny $h_1$}}_{\mbox{\tiny $\pm$}}}
	\newcommand{\Dhtpm}{{\Delta}^{\mbox{\tiny $h_2$}}_{\mbox{\tiny $\pm$}}}

	\newcommand{\dxyz}{d^{\mbox{\tiny $xy$}}_{0}}
	\newcommand{\dxyp}{d^{\mbox{\tiny $xy$}}_{+1}}
	\newcommand{\dxym}{d^{\mbox{\tiny $xy$}}_{-1}}
	\newcommand{\exyp}{e^{\mbox{\tiny $xy$}}_{+1}}
	\newcommand{\exym}{e^{\mbox{\tiny $xy$}}_{-1}}
	\newcommand{\aoz}{a^{0}_{1}}
	\newcommand{\atz}{a^{0}_{2}}
	\newcommand{\aop}{a^{+}_{1}}
	\newcommand{\atp}{a^{+}_{2}}
	\newcommand{\aom}{a^{-}_{1}}
	\newcommand{\atm}{a^{-}_{2}}

	\newcommand{\dzae}{d^{\mbox{\tiny $\mathcal{Z}\mathcal{A}$}}_{\mbox{\tiny $\eta$}}}
	\newcommand{\dzwpme}{d^{\mbox{\tiny $\mathcal{Z}\mathcal{W}^\pm$}}_{\mbox{\tiny $\eta$}}}
	\newcommand{\dawpme}{d^{\mbox{\tiny $\mathcal{A}\mathcal{W}^\pm$}}_{\mbox{\tiny $\eta$}}}
	\newcommand{\dwwe}{d^{\mbox{\tiny $\mathcal{W}\mathcal{W}$}}_{\mbox{\tiny $\eta$}}}
	\newcommand{\ezae}{e^{\mbox{\tiny $\mathcal{Z}\mathcal{A}$}}_{\mbox{\tiny $\eta$}}}
	\newcommand{\ezwpme}{e^{\mbox{\tiny $\mathcal{Z}\mathcal{W}^\pm$}}_{\mbox{\tiny $\eta$}}}
	\newcommand{\eawpme}{e^{\mbox{\tiny $\mathcal{A}\mathcal{W}^\pm$}}_{\mbox{\tiny $\eta$}}}
	\newcommand{\ewwe}{e^{\mbox{\tiny $\mathcal{W}\mathcal{W}$}}_{\mbox{\tiny $\eta$}}}
	\newcommand{\aopm}{a^{\pm}_{1}}
	
	\newcommand{\atpm}{a^{\pm}_{2}}

\newcommand{\sll}{\it}
\newcommand{\itt}{\sl}

\renewcommand{\theequation}{\arabic{section}.\arabic{equation}}

\newcommand{\WW}{{\rm W}}
\newcommand{\F}{{B}}
\newcommand{\A}{{B}}

\newcommand{\mz}{m_{\mbox{\tiny Z}}}
\newcommand{\mh}{m_{\mbox{\tiny H}}}
\newcommand{\mw}{m_{\mbox{\tiny W}}}
\newcommand{\Y}{u}
\newcommand{\om}{\sigma}
\newcommand{\Z}{v}
\newcommand{\W}{v}
\newcommand{\Om}{u}
\newcommand{\f}{f_{1}}
\newcommand{\p}{f_{2}}
\newcommand{\n}{n}

\begin{document}

%\selectlanguage{english}
\title{Stability Analysis of Superconducting Electroweak Vortices}

\author{Julien~Garaud}
\email{garaud@lmpt.univ-tours.fr}
\author{Mikhail~S.~Volkov}
\email{volkov@lmpt.univ-tours.fr}
\
\vspace{1 cm}
\
 \affiliation{ {Laboratoire de Math\'{e}matiques et Physique Th\'{e}orique
CNRS-UMR 6083, \\ Universit\'{e} de Tours,
Parc de Grandmont, 37200 Tours, FRANCE}
}

\begin{abstract}
\vspace{1 cm}

We carry out a detailed stability analysis of the superconducting vortex solutions 
in the Weinberg-Salam theory described in {\iit Nucl.Phys.}~{\bbf B826} (2010) 174.
These vortices are characterized by constant electric current ${\cal I}$
and electric charge density $I_0$, 
for ${\cal I}\to 0$ they reduce to Z strings. 
We consider the generic field fluctuations around the
vortex and apply the functional Jacobi criterion to detect 
the negative modes in the fluctuation operator spectrum. 
We find such modes and determine their dispersion relation,
they turn out to be 
of two different types, according to their spatial behavior.  
There are non-periodic  in space negative modes, which can contribute to the instability of 
infinitely long vortices,
but they can be
 eliminated by imposing the periodic boundary conditions along the vortex. 
There are also periodic negative modes, 
but their wavelength is always larger than a certain minimal value, 
so that they cannot be accommodated by the short 
vortex segments. 
However, even for the latter there remains 
one negative mode  responsible for the homogeneous   
expansion instability. This mode may probably 
be eliminated when the vortex segment is bent into a loop. 
This suggests that small vortex loops  
balanced against contraction by the centrifugal force
could perhaps  be stable.

\end{abstract}

\maketitle
\newpage

\tableofcontents

\section{Introduction}
\setcounter{equation}{0}

Superconducting strings (vortices) were introduced 25 years ago  by Witten
in the context of a simple field theory model containing two 
complex scalars and two Abelian vectors \cite{Witten}. 
They generalize  the well known Abrikosov-Nielsen-Olesen (ANO) vortex \cite{ANO} 
for a non-zero longitudinal current supported by the scalar condensate in the vortex core \cite{SS}. 
The current can be very large, but there is an upper bound for it,  
which is typical 
for the superconductivity models, since too large currents 
produce strong magnetic fields which destroy superconductivity. 
Witten's superconducting strings  have been much studied 
\cite{SS}, \cite{Carter},
mainly in the cosmological context \cite{Vilenkin-Shellard}, \cite{Hindmarsh-Kibble},  
since they can be viewed as solutions of some 
Grand Unification Theory  \cite{Witten} that could perhaps be relevant 
at the early stages of the cosmological evolution. 

The idea that superconducting vortices could also exist
in the Weinberg-Salam theory was suggested longtime ago,
because this theory, 
similar to the Witten model, includes scalar and  
vector fields and admits the `bare' vortices -- Z strings \cite{Zstring}.  
It was therefore    
conjectured that it could also have `dressed Z strings' 
containing a W condensate in the core \cite{per,Olesen}.   
However, when a systematic search of such dressed Z strings gave negative result 
\cite{perk}, the whole idea was abandoned for many years.
Only very recently it was reconsidered again \cite{MV,JGMV2}
and it was found that the negative conclusion  of Ref.~\cite{perk} 
can be circumvented, because it
does not actually forbid the superconducting vortices to exist.
Such solutions have been explicitly constructed in Refs.~\cite{MV,JGMV2}, but 
some of their properties turn out to be quite different as compared to those for the Witten strings.
In particular,  their current can be arbitrarily large,  
since its increase, although quenching the Higgs 
condensate, does not destroy superconductivity, because the current 
is carried by the {vector} W bosons and not by scalars as in Witten's model. 

Superconducting electroweak vortices can be viewed as generalizations 
of Z strings for non-zero electric current and charge. 
They
exist for any value of the 
Higgs boson mass and for any weak mixing angle $\thetaw$. 
Their  current $I_3$ and electric charge density 
$I_0$  transform  as components of a spacelike vector 
$I_\alpha=(I_0,I_3)$ under Lorentz boosts along the vortex. The 
charge can be boosted away 
by passing to the `restframe' where the electric field vanishes, while 
the current never vanishes and can  be defined 
in the Lorentz-invariant way as 
${\cal I}=\sqrt{I_3^2-I_0^2}$. The current is supported by the condensate of  
charged W bosons trapped in the vortex, while outside the vortex the 
massive fields die away and there remains only the Biot-Savart magnetic field
produced by the current. In the ${\cal I}\to 0$ limit the vortices reduce to Z strings.  
For ${\cal I}\gg 1$ they show a large region of size $\sim{\cal I}$
where the Higgs field vanishes, and in the very center of this region there is 
a compact core of size $\sim 1/{\cal I}$ containing the W condensate. 
In this estimates the unit of ${\cal I}$ corresponds to $\sim 10^9$ Amperes, 
so that the current can typically be quite large.
These vortices could perhaps have interesting physical applications, 
but it is important to clarify their stability properties, which is 
the subject of the present paper. 

The fact that Z strings are unstable \cite{Goodband}, \cite{James}, \cite{KO}  
does not necessarily mean  that their superconducting generalizations 
should be unstable too. 
The analysis in the   
semilocal limit, for $\thetaw=\pi/2$, shows that the current-carrying vortices 
do possess instabilities, but
the corresponding negative modes are all inhomogeneous, with the wavelength
always larger than a certain minimal value depending on the current \cite{JGMV}.
 A a result, all instabilities
can be removed by imposing  periodic boundary conditions with a sufficiently
small period.  In this respect the vortex instability is qualitatively similar 
to the hydrodynamical Plateau-Rayleigh  instability of a water 
jet,   or to the Gregory-Laflamme instability of
black strings in the theory of gravity in higher dimensions 
(see \cite{GL} for recent reviews). 

Below we carry out the stability analysis for the  generic superconducting  
electroweak vortices, for $\thetaw<\pi/2$.
We consider the most general field perturbations around the 
vortex and look for negative modes
in the spectrum of the fluctuation operator. Our
main conclusions are as follows. For any values of  current ${\cal I}$
and charge $I_0$ the vortex 
possesses inhomogeneous negative modes which can be periodic or non-periodic 
in space. These instabilities tend, at least as long as the linear perturbation
theory applies, to split the vortex into non-uniform fragments. 
However, imposing the 
periodic boundary conditions with a period $L$ along the vortex  will remove all 
non-periodic negative modes. In addition, if $L$ is small
enough, all inhomogeneous periodic negative modes will be removed as well,
similarly to what  one finds in the semilocal limit \cite{JGMV}. 
Although this suggests  that the periodic vortex segments should be 
stable \cite{JGMV2}, they still possess the {\it homogeneous} perturbation mode, which 
is  not removed  by
periodic boundary conditions, 
since it can be viewed as periodic with any period.  
It is therefore important to know whether this mode is negative or not.
It is actually non-negative for any ${\cal I}$  if $\thetaw=\pi/2$,
and for any $\thetaw$ if ${\cal I}=0$, in which cases 
the periodic vortex segments 
can be stable. 
However, the detailed analysis reveals that 
for generic $\thetaw,{\cal I}$ the homogeneous mode is negative, 
so that the vortices
remain unstable even after imposing the periodic boundary conditions. 
The instability makes them grow thicker. 

At the same time, it is possible that this remaining instability 
can be removed if the vortex segment is bent and its ends are identified to make a loop,
since the thickness of a loop with a fixed radius cannot grow indefinitely. 
It is therefore possible that 
loops made of vortex pieces and balanced against contraction by the 
centrifugal force arising from the momentum circulating along them 
could perhaps be stable. This conjecture can be considered as the `positive'
outcome of our analysis. 
Of course, its verification requires serious efforts,  since one has to 
explicitly construct spinning vortex loops and then study their stability. 
However, any possibility to have stable electroweak solitons can be very important. 

The rest of the paper is organized as follows. In Sec.II 
the electroweak field equations are introduced and their 
vortex solutions are described.   Sec.III considers  the
generic vortex perturbations, separation of variables in the perturbation
equations, gauge fixing, and reduction to a multi-channel Schr\"odinger problem.
Sec.IV describes the Jacobi criterion used to reveal  
the existence of negative modes in the perturbation operator spectrum,
as well as the explicit construction of these modes. The limits of 
zero current and large current are considered,
respectively, in Sec.V and Sec.VI. The electrically charged vortices
are discussed in Sec.VII,
while Sec.VIII contains concluding remarks. 
The two Appendices list the complete equations 
for the background fields and for their perturbations.

\section{Superconducting electroweak vortices}

The bosonic sector of the  
Weinberg-Salam theory is determined by the action density 
\be                             \label{L}
{\cal L}=
-\frac{1}{4g^2}\,\WW^a_{\mu\nu}\WW^{a\mu\nu}
-\frac{1}{4g^{\prime 2}}\,{\F}_{\mu\nu}{\F}^{\mu\nu}
+(D_\mu\Phi)^\dagger D^\mu\Phi
-\frac{\beta}{8}\left(\Phi^\dagger\Phi-1\right)^2. 
\ee
Here the Higgs field $\Phi^{\rm tr}=(\Phi_1,\Phi_2)$
is in the fundamental 
representation of SU(2), its covariant derivative is 
$D_\mu\Phi
=\left(\partial_\mu-\frac{i}{2}\,{\A}_\mu
-\frac{i}{2}\,\tau^a \WW^a_\mu\right)\Phi$ with 
$\tau^a$ being the Pauli matrices, while the field strengths are 
$
\WW^a_{\mu\nu}=\partial_\mu\WW^a_\nu
-\partial_\nu \WW^a_\mu
+\epsilon_{abc}\WW^b_\mu\WW^c_\nu$
and 
$
{\F}_{\mu\nu}=\partial_\mu{\A}_\nu
-\partial_\nu{\A}_\mu$.
The two coupling constants are 
$g=\cos\thetaw$ and
$g^\prime=\sin\thetaw$ where the physical value of the weak mixing angle is
$
\sin^2\thetaw=0.23. 
$
All quantities in \eqref{L} are rendered dimensionless by rescaling,
 their dimensionfull analogues (written in boldface) being 
${\bf \A}_\mu={\mbox{\boldmath $\Phi$}_0}\A_\mu$,  
${\bf W}^a_\mu={\mbox{\boldmath $\Phi$}_0}\WW^a_\mu$,  
${\mbox{\boldmath $\Phi$}}={\mbox{\boldmath $\Phi$}_0}\Phi$, the 
spacetime coordinates ${\bf x}^\mu=x^\mu/{{\bf g}_0\mbox{\boldmath $\Phi$}_0}$.
Here $
\mbox{\boldmath $\Phi$}_0
$
is the Higgs field vacuum expectation value and ${\bf g}_0$ relates to the 
electron charge 
 via 
$
{\bf e}=gg^\prime{\bf \hbar c}\, {\bf g}_0.
$

The theory  is invariant under the  
SU(2)$\times$U(1) gauge transformations
\be                               \label{gauge}
\Phi\to {\rm U}\Phi,~~~~~~~~
{\cal W}\to {\rm U}{\cal W}{\rm U}^{-1}
+2i{\rm U}\partial_\mu {\rm U}^{-1}dx^\mu\,,
\ee
with 
$
{\rm U}=\exp\left(\frac{i}{2}\,\vartheta+\frac{i}{2}\,\tau^a\theta^a\right)
$
where $\vartheta,\theta^a$ are functions of $x^\mu$
and 
$
{\cal W}=
(B_\mu+\tau^a\WW^a_\mu)dx^\mu 
$
is the SU(2)$\times$U(1) Lie-algebra valued gauge field. 
Varying the action with respect to the fields 
gives the field equations,
\begin{align}
\partial^\mu {B}_{\mu\nu}&=g^{\prime 2}\,\frac{i}{2}\,
((D_\nu\Phi)^\dagger\Phi
-
\Phi^\dagger D_\nu\Phi
), \label{P0}\\
D^\mu \WW^a_{\mu\nu}
&=g^{2}\,\frac{i}{2}\,
(
(D_\nu\Phi)^\dagger\tau^a\Phi
-\Phi^\dagger\tau^a D_\nu\Phi
), \label{P1}\\
D_\mu D^\mu\Phi&+\frac{\beta}{4}\,(\Phi^\dagger\Phi-1)\Phi=0,      \label{P2}
\end{align}
with $D_\mu\WW^a_{\alpha\beta}=\partial_\mu \WW^a_{\alpha\beta}
+\epsilon_{abc}\WW^b_\mu\WW^c_{\alpha\beta}$. 

The perturbative mass spectrum of the theory 
contains the photon and the massive Z, W and Higgs 
bosons with masses, respectively, being  
$
\mz={1}/{\sqrt{2}}$, 
$\mw=g\mz$, 
$\mh=\sqrt{\beta}\,\mz$
(in units of 
${\bf e}\mbox{\boldmath $\Phi$}_0/(gg^\prime)$). 
The exact value of the parameter $\beta$ 
is currently unknown, but it is constraint  to belong
 to the 
interval $1.5\leq\beta\leq 3.5$.
Defining  the 
electromagnetic and Z fields as \cite{Nambu}
\be                                  \label{Nambu}
F_{\mu\nu}=\frac{g}{g^\prime}\,  
\A_{\mu\nu}-\frac{g^{\prime}}{g}\,n^a\WW^a_{\mu\nu}\,,~~~~~~
{Z}_{\mu\nu}=\A_{\mu\nu}+n^a\WW^a_{\mu\nu}\,
\ee
with
$
n^a=\Phi^\dagger\tau^a\Phi/(\Phi^\dagger\Phi)
$
 the electromagnetic current density is 
\be                             \label{emcur}
J_\mu=\partial^\nu F_{\nu\mu}.
\ee

A straight vortex oriented along the $x^3$ axis can be described by 
splitting the  spacetime coordinates $x^\mu$ 
into two groups: $x^k=(x^1,x^2)$ spanning 
the 2-planes orthogonal to the vortex, and 
$x^\alpha=(x^0, x^3)$ parameterizing 
the `vortex worldsheet'. Introducing 
the worldsheet vectors
\begin{align}\label{BOOST}
\Sigma_\alpha &=(\sinh(b),\cosh(b))	\, ,~~~~~~~
{\tilde\Sigma}_\alpha  = (\cosh(b),\sinh(b))	\, , ~~~~~~~
\sigma_\alpha=\sigma\Sigma_\alpha,
\end{align}
with $b,\sigma$ being two parameters, 
 one makes the stationary,
cylindrically symmetric field ansatz \cite{JGMV2}
\begin{align}                           \label{003}
{\cal W}&=\Y(\rho)\,\om_\alpha dx^\alpha -
\Z(\rho)\,d\varphi               
+
{\tau}^1\,
[\Om_1(\rho)\,\om_\alpha dx^\alpha - \W_1(\rho)\, d\varphi] \nonumber \\
&+
\tau^3\,
[\Om_3(\rho)\,\om_\alpha dx^\alpha - \W_3(\rho)\, d\varphi],
~~~~~~~~
\Phi=\left(\begin{array}{c}
\f(\rho) \\
\p(\rho)
\end{array}\right),         
\end{align}
where $\f,\p\in\mathbb{R}$ and 
the polar coordinates are introduced, $x^1+ix^2=\rho e^{i\varphi}$.
In what follows 
we shall call $b,\sigma$, respectively,  the boost 
and twist parameters. 
This ansatz keeps its form under Lorentz boosts along the $x^3$ axis
whose only effect is to shift the value of $b$.
This parameter is thus purely cinematic --   
one can always pass to the 
`restframe' where $b=0$ and  the field configuration is purely 
magnetic. 
The ansatz \eqref{003} also keeps its form under gauge 
transformations \eqref{gauge} generated by  
U=$\exp\{-\frac{i}{2}\Gamma \tau^2\}$ with constant $\Gamma$,
whose effect is 
\be                     \label{comp}
(\f+i\p)\to e^{\frac{i}{2}\Gamma}(\f+i\p),~~~~~
(\Om_1+i\Om_3)\to e^{-i\Gamma}(\Om_1+i\Om_3),~~~~
(\W_1+i\W_3)\to e^{-i\Gamma}(\W_1+i\W_3).
\ee

\begin{figure}[ht]
\hbox to \linewidth{ \hss
	\psfrag{y}{}
	\psfrag{lnx}{$\ln(1+\rho)$}
	\psfrag{sigma_u1}{\large $\sigma u_1$}
	\psfrag{sigma_u}{\large$\sigma u$}
	\psfrag{u3}{\large$u_3$}

	\resizebox{8cm}{6cm}{\includegraphics{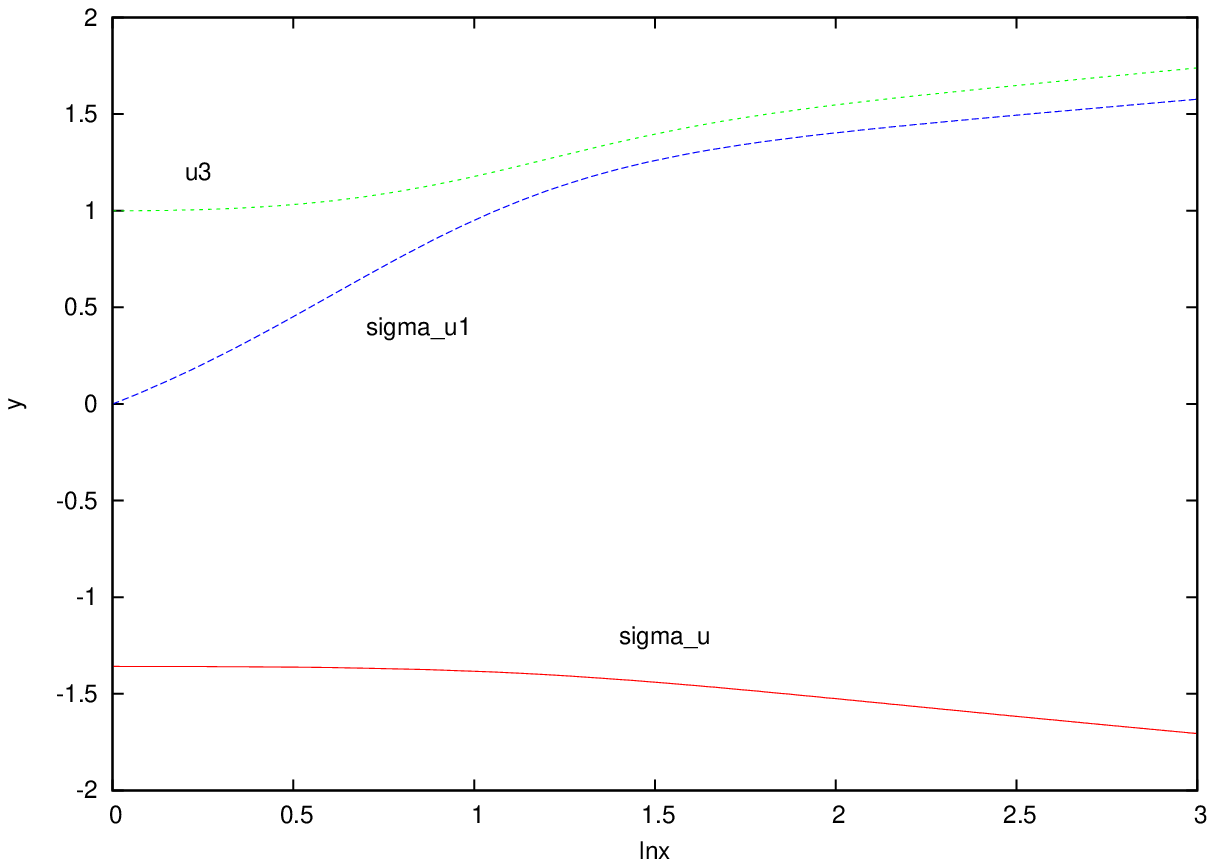}}

\hspace{2mm}
	\psfrag{y}{}
	\psfrag{lnx}{$\ln(1+\rho)$}
	\psfrag{v}{\large$v$}
	\psfrag{v1}{\large$v_1$}
	\psfrag{v3}{\large$v_3$}
	\psfrag{f1}{\large$f_1$}
	\psfrag{f2}{\large$f_2$}
	\resizebox{8cm}{6cm}{\includegraphics{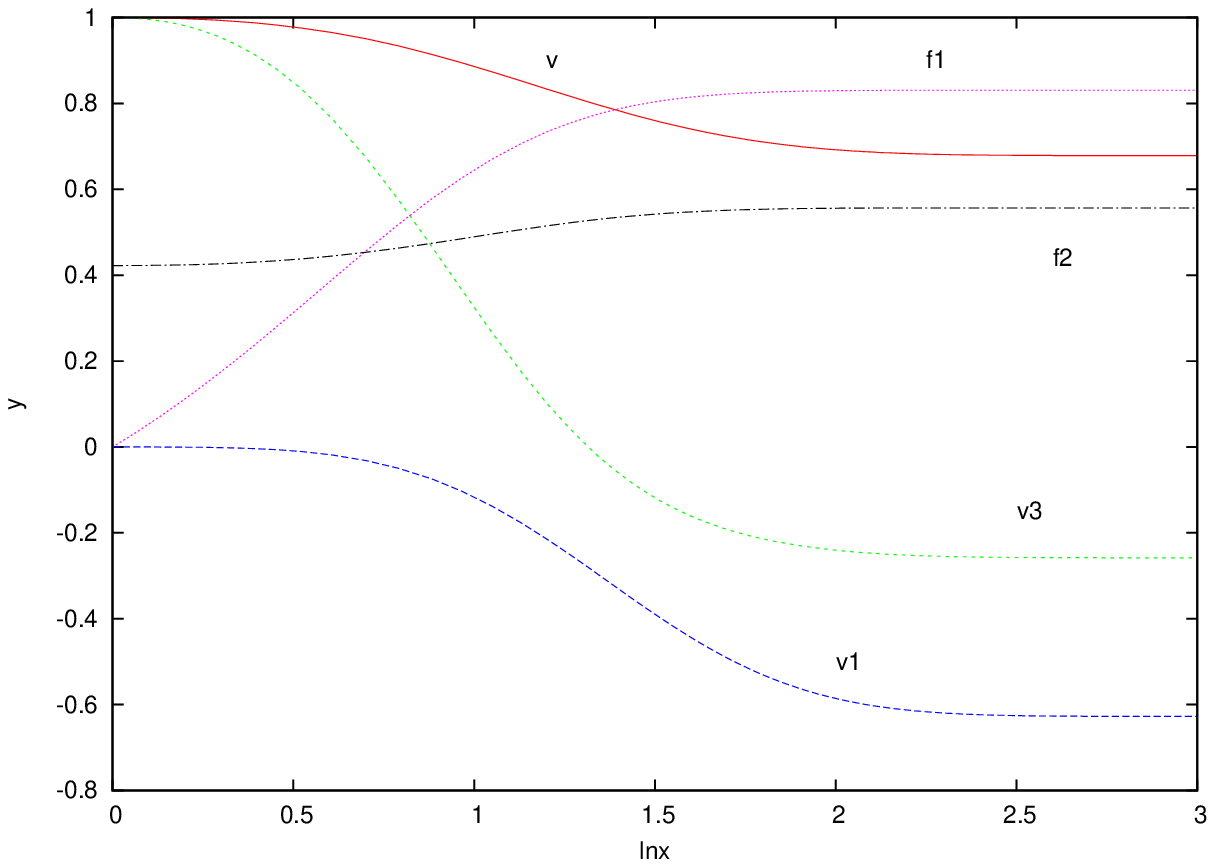}}
\hss}
\caption{Profile functions for the vortex solution with ${\cal I}=2.57$, 
$n=\nu=1$, $\beta=2$, $\sin^2\thetaw=0.23$.
}
\label{Fig2}
\end{figure}

With the parametrization \eqref{003} the field equations 
\eqref{P0}--\eqref{P2} reduce to a system of ordinary 
differential equations \eqref{ee1}--\eqref{CONS1} for the eight functions 
$u,u_1,u_3,v,v_1,v_3,f_1,f_2$ listed in the Appendix A. 
It is worth noting that the boost parameter $b$ drops from these equations,
but they explicitly depend on the twist parameter $\sigma$. 
The boundary conditions for the equations are obtained by 
requiring the energy density to be finite and  
the fields to approach  at large $\rho$ 
the purely electromagnetic  Biot-Savart solution associated with the infinitely 
long electric wire. 
The local analysis in the vicinity of 
$\rho=0,\infty\,$ then gives  the following boundary conditions 
for the field amplitudes for $0\leftarrow \rho\to\infty$ 
(keeping only the leading terms) \cite{JGMV2}
\begin{align}               \label{rec}
a_1\leftarrow\, &\Y\to c_1+Q\ln\rho\,,~~~~~~~~~~~~~~~~~~~~~~~~~~~~~~~~~
2n-\nu\leftarrow\, \Z\to c_2\,,            \nonumber \\  
0\leftarrow\, & \Om_1 \to -(c_1+Q\ln\rho)\sin\gamma\, \,,
~~~~~~~~~~~~~~~~~~~~~~~~~~~0\leftarrow\,  \W_1 \to - c_2\sin\gamma\,,              \nonumber \\ 
1\leftarrow\, & \Om_3 \to -(c_1+Q\ln\rho)\cos\gamma\, \,,~~~~~~~~~~~~~~~~~~~~~~~~~~
\nu\leftarrow\,  \W_3 \to -c_2\cos\gamma\,,              \nonumber \\ 
0\leftarrow\, & \f \to \cos\frac{\gamma}{2}\,, ~~~~~~~~~~~~~~~~~~~~~~~~~~
~~~~~~~~~~~~q\,\delta_n^\nu\leftarrow\,  \p \to \sin\frac{\gamma}{2}\,,
\end{align}
where $a_1,c_1,c_2,Q,\gamma,q$ are real while $n,\nu$ are integers. 
These boundary conditions imply in fact that the vector fields \eqref{003} 
are singular at the symmetry axis, but  this singularity can be removed by 
the gauge transformation \eqref{gauge} 
with 
${\rm U}=e^{i(n-\nu/2)\varphi}e^{i\nu\varphi \tau^3/{2}}$
which renders all fields $\varphi$-dependent, 
\begin{align}                           \label{003a}
{\cal W}&=
\left\{\Y(\rho) +{\tau}_\varphi\, \Om_1(\rho) 
+\tau^3 \Om_3(\rho)\right\}\,\om_\alpha dx^\alpha  \\
&+
\left\{2\n-\nu-\Z(\rho)             
-
{\tau}_\varphi\,\W_1(\rho) 
+\tau^3\, [\nu-\W_3(\rho)]\right\}\, d\varphi,
~~~~ 
\Phi=
\left[\begin{array}{c}
e^{in\varphi}\f(\rho) \\
e^{i(n-\nu)\varphi}\p(\rho)
\end{array}\right]       \nonumber
\end{align}
where $
{\tau}_\varphi=
\tau^1\cos(\nu\varphi  )
 -\tau^2\sin(\nu\varphi )$. 
The logarithmically growing at large $\rho$ terms in the solutions is the specialty  of the 
Biot-Savart field, which is essentially the Coulombian potential in two dimensions.  
The local analysis also shows \cite{JGMV2} that the fields approach their asymptotics 
 \eqref{rec} for $\rho\to\infty$ exponentially fast 
as $e^{-\mh\rho}$, $e^{-\mz\rho}$, $e^{-m_\sigma\rho}$
where $m_\om=\sqrt{\mw^2+\om^2u(\rho)^2}$ is the W-boson mass `dressed' by the interaction 
with the long-range Biot-Savart field. 

Numerically integrating Eqs.\eqref{ee1}--\eqref{ee8}  with the 
boundary conditions \eqref{rec} gives the global 
solutions in the interval $\rho\in[0,\infty)$ \cite{JGMV2}. 
These solutions can be viewed as field-theoretic realizations 
of electric wires, where the wire is represented by a regular 
distribution of massive non-linear fields in the vortex core, 
while in the far field zone 
the massive fields die away and everything reduces to the pure Biot-Savart 
field.

The vortices have the winding number  
$n\geq 1$ and the `polarization' index $\nu=1,2,\ldots\nu_{\rm max}$ 
where $\nu_{\rm max}(\beta,\thetaw,n)$ ranges from $n$ for $\thetaw=\pi/2$ to $2n-1$
for $\thetaw=0$. In addition, they are characterized by  the worldsheet current vector  
\begin{equation} \label{CURRENT}
I_\alpha=\int\partial^\mu  F_{\mu\alpha}\,\dmat^2x
	=-\frac{2\pi Q\sigma_\alpha}{gg^\prime}\equiv {\cal I}\Sigma_\alpha\,.
\end{equation}
Here $I_0={\cal I}\sinh(b)$ is the electric charge per unit vortex length 
and $I_3={\cal I}\cosh(b)$ is the total electric
current through the vortex cross section. 
It is convenient to use ${\cal I},b$ instead of $I_\alpha$ 
 as the solution
parameters.  

The typical profiles of the solutions are shown in Fig.\ref{Fig2}.
When ${\cal I}\to 0$ these
solutions reduce to Z strings, 
that is to the embedded ANO vortices. 
For ${\cal I}\neq 0$
the amplitudes $u,u_1,u_3$ grow with $\rho$ and 
show the logarithmic tails at infinity. 
Solutions up to ${\cal I}\approx 12$ were constructed in Ref.\cite{JGMV2},
and since the dimensionless value ${\cal I}=1$
corresponds to ${\mbox{\boldmath c$\Phi$}_0}=1.8\times 10^9$ Amperes,
the vortex current can in fact be quite large. 
In addition, it seems that
there is no upper bound for possible values of ${\cal I}$, at least in the classical theory.
 This can probably be related to the fact that the vortex current is carried by 
the vector W bosons demonstrating the anti-screening effect \cite{period} --
 the W condensate 
sets up currents which tend to increase the magnetic field and not decrease it
as in the conventional Meissner effect.    
%For large ${\cal I}$ 
%the vortex shows a large region of size $\sim{\cal I}$ where 
%the Higgs field is quenched  to zero by a very strong magnetic field. 
%In the very center of this region there is a compact core of size $\sim1/{\cal I}$ 
%confining  the charged W condensate that carries the current. Outside
%this region the Higgs field relaxes to the vacuum value and everything 
%reduces to the Biot-Savart field. 

When considered in the restframe, where $b=I_0=0$, the vortex is purely magnetic and 
characterized  by its current ${\cal I}$ and  
the magnetic and Z  fluxes.  
Charged (boosted) vortices with $I_0\neq 0$ have in addition  
the electric field, momentum 
and angular momentum.

\section{Perturbing the vortex}
Let us consider small perturbations around 
the vortex configuration $(W^a_\mu,B_\mu,\Phi)$, 
\begin{align}\label{FLUC_1}
&W^a_\mu\to{W^a_\mu}+\delta W^a_\mu	\, ,~~~~~~~~~~~~
B_\mu\to{B_{\mu}}+\delta B_\mu \, ,~~~~~~~~~~ 
\PHI\to{\PHI}+\delta\PHI \, . 
\end{align}
Inserting this into the equations \eqref{P0}--\eqref{P2} and linearizing 
with respect to $\delta W^a_\mu, \delta B_\mu, \delta\PHI$ gives the 
perturbation equations 
\begin{subequations}                \label{LINEA}
\begin{align}
D_\mu D^\mu \delta\PHI &-\imatt\left(\delta B_\mu+\delta W^a_\mu\tau^a\right)D^\mu\PHI
	+\frac{\beta}{4}  \left( 2|\PHI|^2-1 \right)\delta\PHI +\frac{\beta}{4}\delta\PHI^\dagger\PHI^2  \notag \\
	&=\frac{\imatt}{2}\left(\partial_\mu\delta B^\mu
+\tau^a\D_\mu\delta W^{a\mu}\right)\PHI	\, ,\label{LINEA1} \\
\partial_\mu\partial^\mu\delta B^\nu & 
	+\frac{g^{\prime 2}}{2}\left\lbrace \PHI^\dagger\left( \delta B^\nu+\delta W^{a\nu}\tau^a \right)\PHI 
	+2\imatt\left( \delta\PHI^\dagger D^\nu\PHI-\left( D^\nu\PHI \right)\delta\PHI\right) \right\rbrace \notag \\
	 &=  \partial^\nu\left( \partial_\mu\delta B^\mu+
	\frac{\imatt g^{\prime 2}}{2}\left( \delta\PHI^\dagger\PHI -\PHI^\dagger\delta\PHI\right) \right)	\, , \label{LINEA2} \\
\D_\mu\D^\mu \delta W^{a\nu} & +\epsilon_{abc}\delta W^b_\mu W^{c\mu\nu}   \notag \\
	&+\frac{g^2}{2}\left\lbrace \PHI^\dagger\tau^a\delta\PHI\left( \delta B^\nu+\delta W^{c\nu}\delta^a_c \right) 
	+2\imatt\left( \delta\PHI^\dagger \tau^a D^\nu\PHI-\left( D^\nu\PHI \right)\tau^a\delta\PHI\right) \right\rbrace \notag \\
	&=  \D^\nu\left( \D_\mu\delta W^{a\mu}+
	\frac{\imatt g^2}{2}\left( \delta\PHI^\dagger\tau^a\PHI -\PHI^\dagger\tau^a\delta\PHI\right) \right)	\, ,   \label{LINEA3}
\end{align}
\end{subequations} 
with  $\D_\mu  X^a\equiv \partial_\mu X^\mu+\epsilon_{abc}W^b_\mu X^c$.

These equations are invariant under the infinitesimal gauge 
transformations, 
\begin{align} \label{INF_GAUGE_TR}
\delta\PHI &\rightarrow \delta\PHI+\frac{\imatt}{2}\left( \delta\vartheta+\delta\theta^a\tau^a\right)\PHI	\, , 
&\delta B_\mu &\rightarrow \delta B_\mu+ \partial_\mu\delta\vartheta	\, ,
&\delta W^a_\mu &\rightarrow \delta W^a_\mu+ \D_\mu\delta\theta^a	\, .
\end{align}
To suppress  the pure gauge modes, we impose the background gauge conditions, 
\begin{align}\label{BG_GAUGE}
&\partial_\mu\delta B^\mu+
	\frac{\imatt g^{\prime 2}}{2}\left( \delta\PHI^\dagger\PHI -\PHI^\dagger\delta\PHI\right)=0	\, , \notag \\
&\D_\mu\delta W^{a\mu}+
	\frac{\imatt g^2}{2}\left( \delta\PHI^\dagger\tau^a\PHI -\PHI^\dagger\tau^a\delta\PHI\right)=0	\, ,
\end{align}
which eliminates the right hand sides in Eqs.\eqref{LINEA2},\eqref{LINEA3}. 
However, 
this still lives the residual gauge freedom
generated by parameters which fulfill the ghost equations ($n^a$ being defined after \eqref{Nambu})
\begin{align}					\label{ghost}
\partial_\mu\partial^\mu\delta\vartheta&+\frac{g^{\prime 2}}{2}\,\Phi^\dagger\Phi\,
(\delta\vartheta+n^a\delta\theta^a)=0, \notag \\
\D_\mu\D^\mu\delta\theta^a&+\frac{g^{2}}{2}\,\Phi^\dagger\Phi\,
(n^a\delta\vartheta+\delta\theta^a)=0. 
\end{align}

\subsection{Generic perturbation -- separation of variables}

Since the background fields depend only on the radial coordinate $\rho$, 
we can Fourier-decompose the perturbations with respect to $x^\alpha,\varphi$.  
Keeping in mind the action of the Lorentz  boosts on the 
background solutions, we wish to keep track of their action 
on the perturbations too.   We therefore introduce 
$\phase\equiv (\omega{\tilde\Sigma}_\alpha+\kappa\Sigma_\alpha)x^\alpha+m\varphi$,
which reduces in the restframe to $\omega x^0+\kappa x^3+m\varphi$.
Denoting $\delta W^0_\mu\equiv \delta B_\mu $ the 
generic perturbations can be decomposed as 
\begin{align}\label{DECOMP}
\delta \Phi_{\rm a} &=\sum_\okm \left\{[\phi_{\rm a}(\okm|\rho)
+i\,\psi_{\rm a}(\okm|\rho)]\cos\phase \, \right. \notag \\
 &\left. + [\pi_{\rm a}(\okm|\rho)+
\imatt\,\chi_{\rm a}(\okm|\rho)]\sin\phase \right\}	\, , \notag \\
-\delta W^a_\mu \tilde{\Sigma}^\mu &= 
	\sum_\okm \left\{ X^a_1(\okm|\rho)\cos\phase + 
Y^a_1(\okm|\rho)\sin\phase\right\} \, ,\notag \\
-\delta W^a_\mu \Sigma^\mu &= 
	\sum_\okm \left\{X^a_4(\okm|\rho)\cos\phase 
+ Y^a_4(\okm|\rho)\sin\phase\right\} \, ,\notag \\ 
\delta W^a_k &= \sum_\okm \left\{X^a_k(\okm,\rho)\cos\phase 
+Y^a_k(\okm,\rho)\sin\phase\right\} \, ,
\end{align}
where ${\rm a}=1,2$ and $k=1,2$ while now $a=0,1,2,3$.  The infinitesimal gauge 
transformations Eq.\eqref{INF_GAUGE_TR} can be decomposed in the same way (with
$\theta^0\equiv \vartheta$) 
\begin{align}\label{DECOMP_GT}
\delta\theta^a &= \sum_\okm \left\{\alpha^a
(\okm|\rho)\cos\phase+\gamma^a(\okm|\rho)\sin\phase\right\} \, .
\end{align}
Inserting the decompositions \eqref{DECOMP} into Eqs.\eqref{LINEA} 
the variables $x^\alpha,\varphi$ decouple and one obtains, for given
$\okm$, 
a system of $40$ ordinary differential equations for the 40 radial functions 
$\phi_{\rm a},\ldots ,Y^a_k$ in \eqref{DECOMP}. 
These $40$ equations split (if $\kappa\in\mathbb{R}$) 
into $2$ independent subsystem of $20$ equations each.
These subsystems are identical to each other 
upon the replacement 
\begin{align}\label{REPLACE}
\pi_{\rm a} &\leftrightarrow \phi_{\rm a}	\, , &
\psi_{\rm a} &\leftrightarrow -\chi_{\rm a}	\, , \notag \\
Y^a_k &\leftrightarrow X^a_k	\, , &
Y^2_2 &\leftrightarrow X^2_2	\, ,\notag\\
X^a_2 &\leftrightarrow -Y^a_2	\, , &
X^2_k &\leftrightarrow -Y^2_k	\, . 
\end{align}
Here ${\rm a}=1,2$ but $a=0,1,3$ and $k=1,3,4$ (where possible we shall not write
explicitly  the arguments $(\okm|\rho)$). 
Such a splitting of 
the equations into two groups is the consequence of the fact that the 
background configurations \eqref{003} are {\it real},  and so that  
the real and imaginary parts of their perturbations should be independent. 
In Sec.\ref{cvort} below we shall study the case of complex $\kappa,\omega$, and then the 
40 equations do not split into two subsystems any more, but for the time being $\kappa$ is real and  
we can restrict our analysis to the 20 equations.
These are equations for the 20 radial amplitudes
in the right hand sides of \eqref{REPLACE},  
they factorize with $\cos\phase$.

These equations are rather long and we do not write them down explicitly.  
Not all of them are independent, since there are four 
identities relating them to each other. 
These are the linearized versions of the identities 
obtained by taking the divergences of 
the vector field equations \eqref{P0} and \eqref{P1}, 
their existence is the manifestation of the gauge invariance. 
The equations are also invariant under the action of the gauge 
transformations, which now assume the following explicit form: 
\begin{align}                           \label{g-rad}
 X^0_1 &\rightarrow X^0_1-\omega\ \gamma^0 \, ,&
 X^1_1 &\rightarrow X^1_1-\omega\ \gamma^1 \notag \, ,\\
 Y^0_2 &\rightarrow Y^0_2+\ (\gamma^0)^\prime \, ,&
 Y^1_2 &\rightarrow Y^1_2+\ (\gamma^1)^\prime \notag \, ,\\
 X^0_3 &\rightarrow X^0_3+m\ \gamma^0 \, ,&
 X^1_3 &\rightarrow X^1_3+m\ \gamma^1 +v_3\alpha^2 \notag \, ,\\
 X^0_4 &\rightarrow X^0_4+\kappa\ \gamma^0 \, ,&
 X^1_4 &\rightarrow X^1_4+\kappa\ \gamma^1 -\sigma u_3\alpha^2 \notag \, ,\\
  Y^2_1 &\rightarrow Y^2_1+\omega\ \alpha^2  \, , &
X^3_1 &\rightarrow X^3_1-\omega\ \gamma^3  \notag \, ,\\
  X^2_2 &\rightarrow X^2_2+\ (\alpha^2)^\prime\, ,  &
 Y^3_2 &\rightarrow Y^3_2+\ (\gamma^3)^\prime \notag\, , \\
 Y^2_3 &\rightarrow Y^2_3-m\ \alpha^2 +(v_1\gamma^3-v_3\gamma^1) \, , &
 X^3_3 &\rightarrow X^3_3+m\ \gamma^3 -v_1\alpha^2 \notag  \, ,\\
 Y^2_4 &\rightarrow Y^2_4-\kappa\ \alpha^2 +\sigma(u_3\gamma^1-u_1\gamma^3) \, , &
  X^3_4 &\rightarrow X^3_4+\kappa\ \gamma^3 +\sigma u_1\alpha^2\notag \, ,  \\
 \phi_1 &\rightarrow \phi_1+\frac{1}{2}\,\alpha^2f_2  \, , &
 \chi_1 &\rightarrow \chi_1+
 		\frac{1}{2}[( \gamma^0+\gamma^3)f_1+ \gamma^1f_2] \notag  \, , \\
 \phi_2 &\rightarrow  \phi_2-\frac{1}{2}\,\alpha^2f_1  \, , &
\chi_2 &\rightarrow \chi_2+		\frac{1}{2}[\gamma^1f_1+( \gamma^0-\gamma^3)f_2], 
\end{align}
where the prime denotes differentiation with respect to $\rho$. 

\subsection{Gauge fixing}

The additional terms on the right in \eqref{g-rad} are pure gauge modes. 
They automatically fulfill the perturbation equations for any gauge functions 
$\gamma^0\equiv\gamma^0(\okm|\rho),\gamma^1,\alpha^2,\gamma^3$
(verification of this is a good consistency check). We need to impose gauge
conditions to eliminate  these non-physical solutions. For example, one can use  
the temporal gauge, which completely eliminates all  gauge degrees of freedom \cite{JGMV}. 
However, the fluctuation operator becomes then rather complicated.
We have therefore chosen to use the background gauge conditions \eqref{BG_GAUGE}, they 
lead to more easy to handle equations, although  
not  eliminating completely   all  gauge modes. 

After separating the variables the background gauge conditions 
\eqref{BG_GAUGE} reduce to four constraint equations 
\begin{align} \label{GAUGE_CONSTR}
\omega & X^0_1 - \left( \partial_\rho+\frac{1}{\rho}\right) Y^0_2 +\frac{m}{\rho^2} X^0_3
	+\kappa X^0_4 +g^{\prime 2}\left( f_1\chi_1+ f_2\chi_2 \right)=0	\, , \notag  \\
\omega & X^1_1 - \left( \partial_\rho+\frac{1}{\rho}\right) Y^1_2 +\frac{m}{\rho^2}X^1_3
	+\kappa X^1_4 +g^2\left( f_2\chi_1+f_1\chi_2 \right) +\sigma u_3 Y^2_4 -\frac{v_3}{\rho^2}Y^2_3=0	\, , \notag \\
-\omega & Y^2_1 - \left( \partial_\rho+\frac{1}{\rho}\right) X^2_2 -\frac{m}{\rho^2}Y^2_3
	-\kappa Y^2_4 +g^2\left( f_2\phi_1-f_1\phi_2 \right)  \notag \\
	& +\sigma \left(u_1X^3_4-u_3X^1_4  \right)	+\frac{1}{\rho^2}\left(v_3X^1_3-v_1X^3_3\right)=0	\, , \notag \\
\omega & X^3_1 - \left( \partial_\rho+\frac{1}{\rho}\right) Y^3_2 +\frac{m}{\rho^2}X^3_3
	+\kappa X^3_4 +g^2\left( f_1\chi_1-f_2\chi_2 \right)  -\sigma u_1 Y^2_4 +\frac{v_1}{\rho^2}Y^2_3=0	\, .
\end{align} 
Imposing these, one discovers that the 20 radial equations split into two independent subsystems as 4+16,
since the four amplitudes in \eqref{GAUGE_CONSTR} which are 
proportional to $\omega$ 
decouple from the remaining 16 amplitudes. 
Let us call these four amplitudes {temporal}, 
they are governed by the  equations 
\begin{equation}\label{TEMPORAL}
\left( \begin{array}{cccc} 
D_1	&S	&0	&T	\\	
S	&D_2	&U	&W	\\	
0	&U	&D_3	&V	\\	
T	&W	&V	&D_4	\end{array}\right)
\left( \begin{array}{c}
	X^0_1/g^\prime \\
	X^1_1/g \\
	Y^2_1/g \\
	X^3_1/g 
       \end{array}\right)=0	\, , 
\end{equation}
where
\begin{align}
D_1 &= -\diff+\frac{m^2}{\rho^2}+\kappa^2-\omega^2+\frac{g^{\prime 2}}{2}\left(f_1^2+f_2^2\right)	\, ,\notag \\
D_2 &= -\diff+\frac{m^2+v_3^2}{\rho^2}\sigma^2u_3^2+\kappa^2-\omega^2+\frac{g^2}{2}\left(f_1^2+f_2^2\right)	\, ,\notag \\
D_3 &= -\diff+\frac{m^2+v_1^2+v_3^2}{\rho^2}\sigma^2(u_1^2+u_3^2)+\kappa^2-\omega^2+\frac{g^2}{2}\left(f_1^2+f_2^2\right)	\, ,\notag \\
D_4 &= -\diff+\frac{m^2+v_1^2}{\rho^2}\sigma^2u_1^2+\kappa^2-\omega^2+\frac{g^2}{2}\left(f_1^2+f_2^2\right)	\, ,
\end{align}
and the off-diagonal terms are 
\begin{align}
S &= gg^\prime f_1f_2 \, ,&
T &= gg^\prime (f_1^2-f_2^2) \, ,\notag \\
U &= -2\left(\frac{mv_3}{\rho^2}+\kappa\sigma u_3\right) \, ,&
V &= -2\left(\frac{mv_1}{\rho^2}-\kappa\sigma u_1\right) \, , &
W &= -\left(\frac{v_1v_3}{\rho^2}+\sigma^2u_1u_3 \right) \, .
\end{align}
A direct verification reveals that if one resolves the constraints \eqref{GAUGE_CONSTR}
with respect to the temporal amplitudes, then the temporal equations 
will be automatically fulfilled by virtue of the equations for the remaining 16 amplitudes. 
The latter are described by Eqs.\eqref{SCHRODINGER} below. 
Every solution of the $16$-channel problem \eqref{SCHRODINGER}
therefore generates a solution of the temporal equations \eqref{TEMPORAL}.
This can be understood by noting that the temporal equations  coincide with the 
ghost equations.

The ghost equations describe the residual gauge freedom left in the background gauge.  
They can be obtained by 
inserting the pure gauge modes in \eqref{g-rad} into \eqref{GAUGE_CONSTR}, or equivalently 
injecting the mode decomposition 
\eqref{DECOMP_GT} into \eqref{ghost}. This gives 
four radial equations for the  gauge parameters $\gamma^0,\gamma^1,\alpha^2,\gamma^3$
which  coincide with the temporal equations
\eqref{TEMPORAL} upon the replacement 
\begin{align}
X^0_1 &\leftrightarrow \gamma^0 \, , &
X^1_1 &\leftrightarrow \gamma^1 \, , &
Y^2_1 &\leftrightarrow -\alpha^2 \, , &
X^3_1 &\leftrightarrow \gamma^3 \, . 
\end{align}
Therefore, the temporal amplitudes are pure gauge modes. It follows that they 
can be constructed via resolving the constraints \eqref{GAUGE_CONSTR} if only the 
corresponding solutions of \eqref{SCHRODINGER} are also pure gauge. 
Resolving the constraints for a non-pure gauge solution of \eqref{SCHRODINGER}
should also give a solution of the temporal equations \eqref{TEMPORAL}, but since it cannot then be 
pure gauge, it can only be trivial. This gives a simple recipe to 
distinguish between the physical and unphysical solutions of the $16$-channel Schr\"odinger system
\eqref{SCHRODINGER}: if a solution fulfills the constraints \eqref{GAUGE_CONSTR} with zero temporal amplitudes 
then it is non-trivial, otherwise it is pure gauge.

We have explicitly tested this recipe for the negative modes of the system \eqref{SCHRODINGER}. 
These modes are all physical, since the spectrum of the 
ghost operator is positive, 
and because they fulfill the constraints \eqref{GAUGE_CONSTR} 
with $X^0_1=X^1_1=Y^2_1=X^3_1=0$. 

Since the four temporal amplitudes vanish for the physical solutions, one can use the constraints 
\eqref{GAUGE_CONSTR} in order to algebraically express four other amplitudes (for example those 
proportional to $\kappa$) in terms of the remaining 12 amplitides. The system 
\eqref{SCHRODINGER} then reduces to 12 independent equations only,
which coincide with  the equations obtained in the temporal gauge.  
However, their structure turns out to be rather complicated,  
which is why we prefer to work with the 16-channel system \eqref{SCHRODINGER}.

\subsection{Reduction to a Schr\"odinger problem}

Imposing the background gauge conditions decouples  the 4 
temporal/ghost amplitudes, 
while the equations for the remaining 16 amplitudes   
can be cast into a Schr\"odinger form after the following operations. 
 We redefine the amplitudes as 
\begin{align}
Y^0_2&= \frac{g^\prime}{\sqrt{2}}\left(\frac{g^\prime}{g}\left(\Zp +\Zm \right)+\Ap+\Am \right)\, , 
	& Y^3_2&=\frac{1}{\sqrt{2}}\left(g\left(\Zp +\Zm\right)-g^\prime\left(\Ap+\Am\right) \right) \, , \notag \\
X^0_3&= g^\prime\frac{\rho}{\sqrt{2}}\left(\frac{g^\prime}{g}\left(\Zp -\Zm \right)+\Ap-\Am \right)\, , 
	& X^3_3&=\frac{\rho}{\sqrt{2}}\left(g\left(\Zp -\Zm\right)-g^\prime\left(\Ap-\Am\right) \right) \, , \notag \\
X^0_4&= g^\prime\left(\frac{g^\prime}{g}\Zz+\Az \right)\, , 
	& X^3_4&=\left(g\Zz-g^\prime\Az \right) \, , \notag 
\end{align}
\begin{align}       \label{lincomb}
Y^1_2&=\frac{1}{2} \left(\Wpp+\Wpm+\Wmp+\Wmm \right) \, , 
	& X^2_2&=\frac{1}{2}\left(\Wpp+\Wpm-\Wmp-\Wmm \right) \, , \notag \\
X^1_3&=\frac{\rho}{2}\left(\Wpp-\Wpm+\Wmp-\Wmm \right) \, , 
	& Y^2_3&=\frac{\rho}{2}\left(-\Wpp+\Wpm+\Wmp-\Wmm \right) \, , \notag \\
X^1_4&= \frac{1}{\sqrt{2}}\left( \Wmz+\Wpz\right) \, , 
	& Y^2_4&= \frac{1}{\sqrt{2}}\left( \Wmz-\Wpz\right) \, , \notag \\
& & & \notag \\
\phi_1 &=\frac{1}{2g}\left( \Hom-\Hop\right) \, , 
	& \chi_1&=\frac{1}{2g}\left( \Hom+\Hop\right)\, , \notag \\
\phi_2 &=\frac{1}{2g}\left( \Htm-\Htp\right) \, , 
	& \chi_2&=\frac{1}{2g}\left( \Htm+\Htp\right)\, . 
\end{align}
Here the notation $\mathcal{A}$, $\mathcal{Z}$ and $\mathcal{W}^\pm$ 
reflect the fact that these amplitudes correspond to 
the photon, Z and W bosons, respectively.
The subscripts refer to their polarizations. 
Introducing  the 16-component vector 
\be
\Psi^{\rm tr}=\left(\Zz,\Zp,\Zm,\Az,\Ap,\Am,\Wpz,\Wpp,\Wpm,\Wmz,\Wmp,
\Wmm,\Hop,\Hom,\Htp,\Htm\right)\,,
\ee
the  equations assume the form  
\be \label{SCHRODINGER}
-\frac{1}{\rho}\left(\rho\Psi^\prime\right)^\prime+
\mathcal{U}(\kappa,m|\rho)\Psi=\omega^2\Psi	\, , 
\ee
where 
$\mathcal{U}$ is a $16\times16$ symmetric potential energy matrix depending on the background 
fields. Its explicit form  is given in the Appendix B. 
These equations are
invariant under 
$\omega\to-\omega$, $\kappa\to -\kappa$ and $m\to-m$ provided that 
\begin{align}\label{TRANS_PARAM}
\Zz(\okm|\rho) &\to -\Zz(-\omega,-\kappa,-m|\rho)	\, ,&
\Zpm(\okm|\rho)&\to \Zmp(-\omega,-\kappa,-m|\rho)	\, , \notag\\
\Az(\okm|\rho)&\to -\Az(-\omega,-\kappa,-m|\rho)	\, ,&
\Apm(\okm|\rho)&\to \Amp(-\omega,-\kappa,-m|\rho)	\, , \notag\\
\Wpmz(\okm|\rho)&\to -\Wmpz(-\omega,-\kappa,-m|\rho)	\, ,&
\Wpmpm(\okm|\rho)&\to \Wmpmp(-\omega,-\kappa,-m|\rho)	\, ,\notag\\
h_{\rm a}^\pm(\okm|\rho)&\to h_{\rm a}^\mp(-\omega,-\kappa,-m|\rho) 	\, .
\end{align}

\subsection{Boundary conditions}

The small $\rho$ behavior of the perturbations can be determined by solving Eqs.\eqref{SCHRODINGER} 
in power series. 
For each of the 16 equations we find two solutions, 
one of which is bounded for $\rho\to 0$
while the other one is divergent.  
The bounded solutions are 
\begin{align}\label{BC_AXIS}
\Ze&=c_\eta^{\mbox{\tiny $Z$}}\rho^{|m-\eta|}+\dots	\, , &
	\Ae&=c_\eta^{\mbox{\tiny $A$}}\rho^{|m-\eta|}+\dots	\, , &
	\Wpme&=c_\eta^{\mbox{\tiny $W^\pm$}}\rho^{|\nu\pm(m-\eta)|}+\dots	\, , \notag \\
\Hopm&=c_\eta^{\mbox{\tiny $h_1^\pm$}}\rho^{|n\mp m|}+\dots	\, , & 
	\Htpm&=c_\eta^{\mbox{\tiny $h_2^\pm$}}\rho^{|n-\nu\mp m|}+\dots	\, ,
\end{align}
where $c_\eta^{\mbox{\tiny $Z$}}$, $c_\eta^{\mbox{\tiny $A$}}$, 
$c_\eta^{\mbox{\tiny $W^\pm$}}$, $c_\eta^{\mbox{\tiny $h_a^\pm$}}$
are $16$ integration constants and the dots stand for subleading terms.

%\subsubsection{Boundary conditions at infinity}

We are interested in bound state type solutions  for which $\Psi\to 0$
as $\rho\to\infty$. 
In order to work out their  behavior  at large $\rho$, 
it is convenient to temporarily pass to the gauge where $f_2(\infty)=0$. 
This is achieved by applying the global symmetry \eqref{comp} with $\Gamma=-\gamma$, 
which 
corresponds to the gauge transformation 
\eqref{gauge} with U$=\exp\{\frac{i}{2}\gamma\}$.
The background fields then simplify and one finds at large $\rho$ 
\begin{align}\label{BC_ASYMPT}
\Ze &= \frac{b_\eta^{\mbox{\tiny $Z$}}}{\sqrt{\rho}}\EXP{-\mu_{\mbox{\tiny $Z$}}\rho}+\dots	\, ,~~~~~~~~~
	\Ae ~= \frac{b_\eta^{\mbox{\tiny $A$}}}{\sqrt{\rho}}\EXP{-\mu_{\mbox{\tiny $A$}}\rho}+\dots	\, ,~~~~~~
	\Hop+\Hom = \frac{b_+^{\mbox{\tiny $h_1$}}}{\sqrt{\rho}}\EXP{-\mu_{\mbox{\tiny $Z$}}\rho}+\dots	\, ,	\notag\\
\Wpme &= \frac{b_\eta^{\mbox{\tiny $W^\pm$}}}{\sqrt{\rho}}\EXP{-\int\mu_{\mbox{\tiny $W_\pm$}}d\rho}+\dots	\, ,~
	\Htpm = \frac{b_\pm^{\mbox{\tiny $h_2$}}}{\sqrt{\rho}}\EXP{-\int\mu_{\mbox{\tiny $W_\pm$}}d\rho}+\dots	\, ,~
	\Hop-\Hom = \frac{b_-^{\mbox{\tiny $h_1$}}}{\sqrt{\rho}}\EXP{-\mu_{\mbox{\tiny $H$}}\rho}+\dots	\, .	
\end{align}
Here the {effective} mass terms  
\begin{align}\label{MASSES}
\mu^2_{\mbox{\tiny $A$}} &= \kappa^2-\omega^2	\, , &
\mu^2_{\mbox{\tiny $Z$}} &= \mu^2_{\mbox{\tiny $A$}}+\mz^2	\, ,&
\mu^2_{\mbox{\tiny $H$}} &= \mu^2_{\mbox{\tiny $A$}}+\mh^2	\, , &
\mu^2_{\mbox{\tiny $W_\pm$}}(\rho) &= 
\left(\sigma u(\rho)\pm \kappa \right)^2-\omega^2+\mw^2	\, &
\end{align}
are assumed to be positive and  
$b_\eta^{\mbox{\tiny $Z$}}$, $b_\eta^{\mbox{\tiny $A$}}$, 
$b_\eta^{\mbox{\tiny $W^\pm$}}$, $b_\eta^{\mbox{\tiny $h_a^\pm$}}$
are $16$ integration constants while the dots stand for the subleading terms. 
One can now apply  to the whole system (background + perturbations) 
the inverse gauge rotation with U$=\exp\{-\frac{i}{2}\gamma\}$. 
The background then returns to the gauge 
where $f_2(\infty)=\sin\frac{\gamma}{2}$ while the perturbations 
\eqref{BC_ASYMPT} change as
\begin{align}\label{GLOBAL}
 \Ze~ &\to \left(g^{\prime 2}+g^2\cos\gamma\right)\Ze+2gg^\prime\sin^2\frac{\gamma}{2}\Ae
	-\frac{g}{\sqrt{2}}\Wpe\sin\gamma-\frac{g}{\sqrt{2}}\Wme\sin\gamma	\, ,\notag \\
 \Ae~ &\to \left(g^2+g^{\prime 2}\cos\gamma\right)\Ae+2gg^\prime\sin^2\frac{\gamma}{2}\Ze
	+\frac{g^\prime}{\sqrt{2}}\Wpe\sin\gamma+\frac{g^\prime}{\sqrt{2}}\Wme\sin\gamma	\, ,\notag \\
 \Wpe &\to \Wpe\cos^2\frac{\gamma}{2}-\Wme\sin^2\frac{\gamma}{2}
	+\frac{g}{\sqrt{2}}\Ze\sin\gamma-\frac{g^\prime}{\sqrt{2}}\Ae\sin\gamma	\, ,\notag \\
 \Wme &\to \Wme\cos^2\frac{\gamma}{2}-\Wpe\sin^2\frac{\gamma}{2}
	+\frac{g}{\sqrt{2}}\Ze\sin\gamma-\frac{g^\prime}{\sqrt{2}}\Ae\sin\gamma	\, ,\notag \\
 \Hopm &\to \Hopm\cos\frac{\gamma}{2}-\Htpm\sin\frac{\gamma}{2}	\, ,\, \, \,
 \Htpm \to \Htpm\cos\frac{\gamma}{2}+\Hopm\sin\frac{\gamma}{2}	\, .
\end{align}
This gives the large $\rho$ behavior of perturbations. 
At this point 
we have everything we need to solve the perturbation equations \eqref{SCHRODINGER}.

\section{Stability test}

Summarizing the above analysis, we have arrived at the eigenvalue problem \eqref{SCHRODINGER} and now
we wish to know whether it admits bound state solutions with $\omega^2<0$. If exist, such solutions
would correspond to unstable modes of the background vortex.
In order to detect them, one  possibility is to
directly integrate the 16 coupled second order differential equations \eqref{SCHRODINGER}.
However, if one just wants  to know if negative modes exist or not, 
it is not necessary to construct them explicitly. 
A simple method to reveal their existence is to use the Jacobi criterion \cite{GELFAND},
which essentially uses the fact that 
the ground state wave function does not oscillate while the excited states do.   
 It follows that if the zero
energy wave function oscillates then the ground state energy 
is negative. 

\subsection{Jacobi criterion}

When applied to our problem the Jacobi method gives the
following recipe.
 Let $\Psi_s(\rho)$ with $s=1,\dots,16$ be the 
$16$ linearly independent, regular at the symmetry axis solutions of \eqref{SCHRODINGER}. Each 
of them is a $16$-component vector, 
$\Psi_s(\rho)\equiv \Psi^I_s(\rho)$, $I=1,\dots,16$. Let $\Delta(\rho)$ be the determinant
of the matrix $\Psi^I_s(\rho)$. If it vanishes somewhere, then there exists 
a negative part of the spectrum. According to \cite{AMMAN}, the number of zeros of $\Delta(\rho)$ 
is equal to the number of negative modes. 

\begin{figure}[ht]
\hbox to \linewidth{ \hss
	\psfrag{y}{}
	\psfrag{lnx}{$\ln(1+\rho)$}
	\psfrag{k>2s}{\large $\kappa>\kappa_{\mbox{\tiny max}}$}
	\psfrag{0<=k<=2s}{\large $0\leq\kappa\leq\kappa_{\mbox{\tiny max}}$}

	\psfrag{k>=1.4}{\large $\kappa>\kappa_{\mbox{\tiny max}}$}
	\psfrag{0<=k<1.4}{\large $0\leq\kappa\leq\kappa_{\mbox{\tiny max}}$}

	\resizebox{8cm}{6cm}{\includegraphics{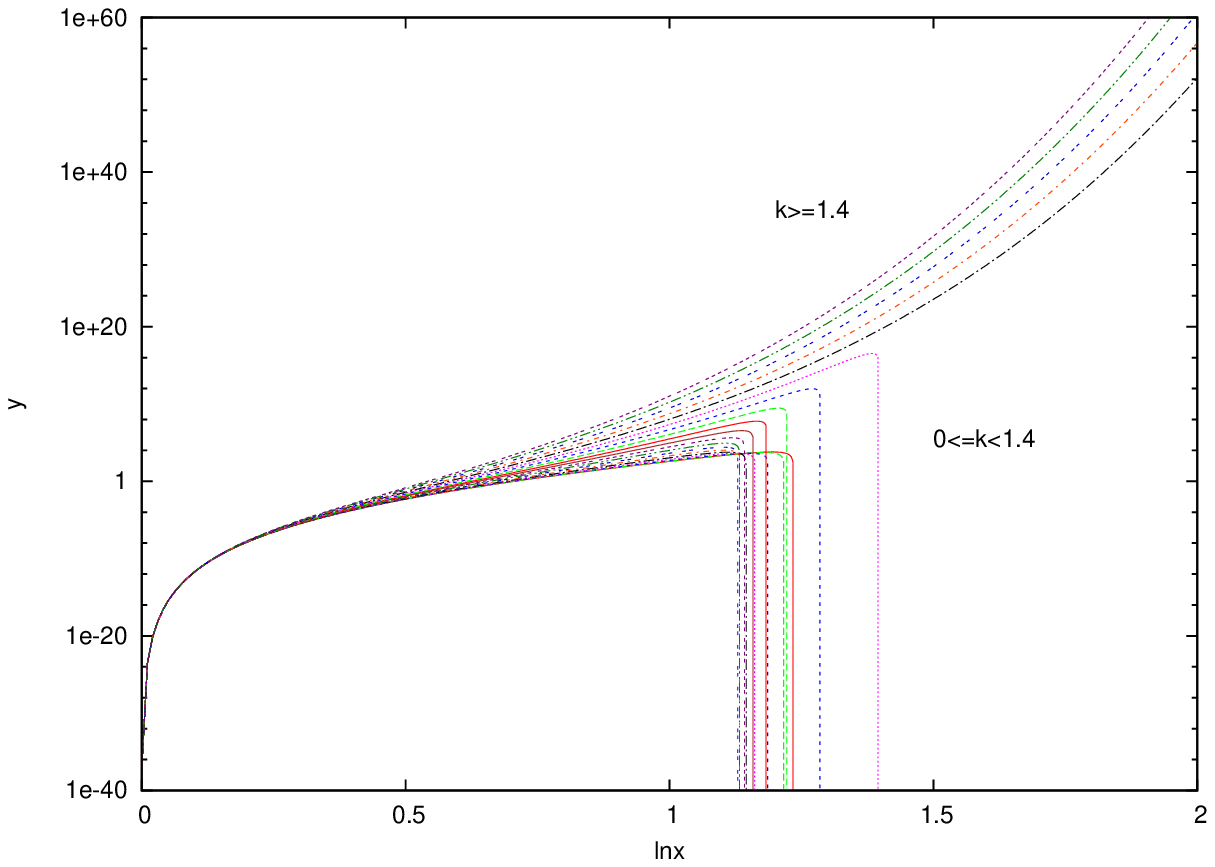}}

\hspace{2mm}

	\resizebox{8cm}{6cm}{\includegraphics{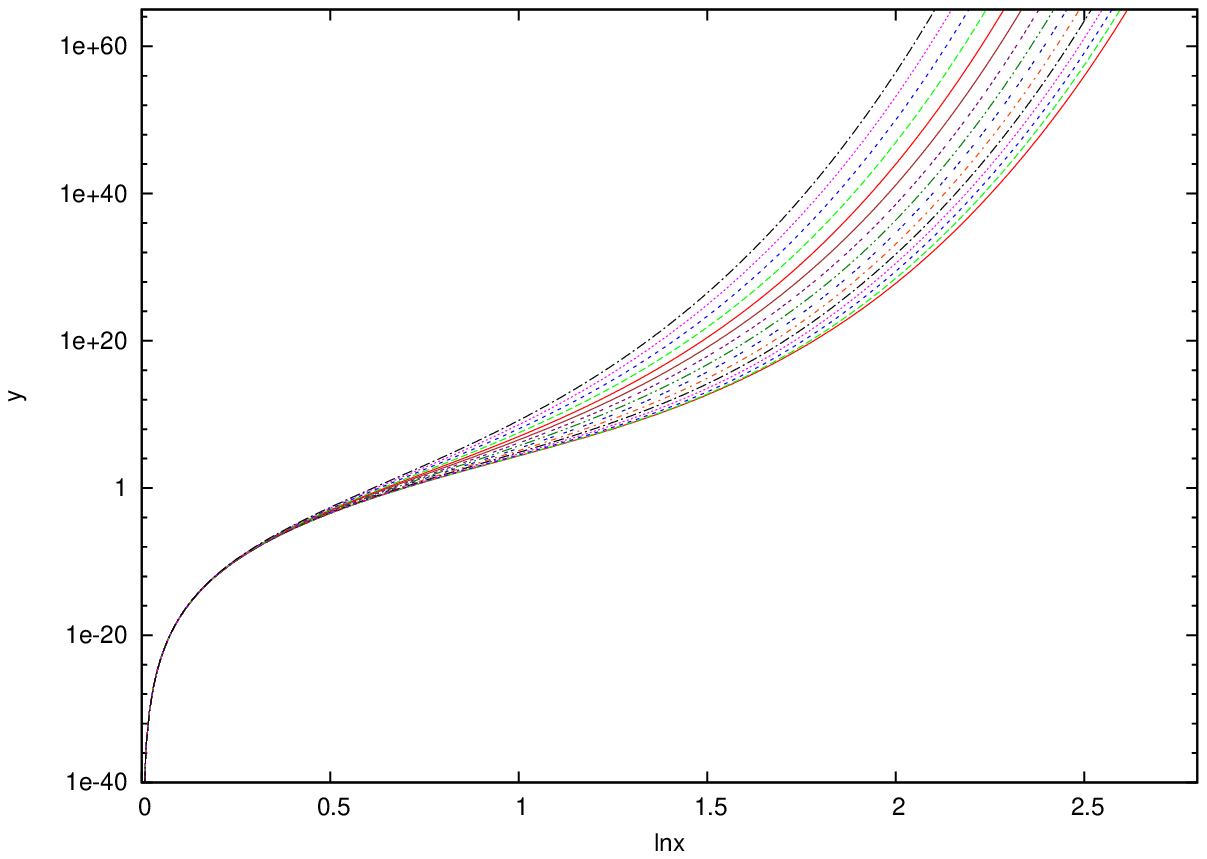}}
\hss}
\caption{The behavior of the Jacobi determinant $\Delta(\rho)$ 
for  fluctuations  around the 
$n=\nu=1$, ${\cal I}=0.87$ vortex 
($\beta=2$, $\sin^2\thetaw=0.23$) 
for different values of $\kappa$ for
$m=0$ (left) and for  $m=1$ (right). The behavior for $m=2$
is qualitatively the same as for $m=1$.}
\label{Fig_n=1}
\end{figure}

\begin{figure}[ht]
\hbox to \linewidth{ \hss
	\psfrag{y}{}
	\psfrag{lnx}{$\ln(1+\rho)$}
	\psfrag{k>2s}{\large $\kappa>\kappa_{\mbox{\tiny max}}$}
	\psfrag{0<=k<=2s}{\large $0\leq\kappa\leq\kappa_{\mbox{\tiny max}}$}

	\resizebox{8cm}{6cm}{\includegraphics{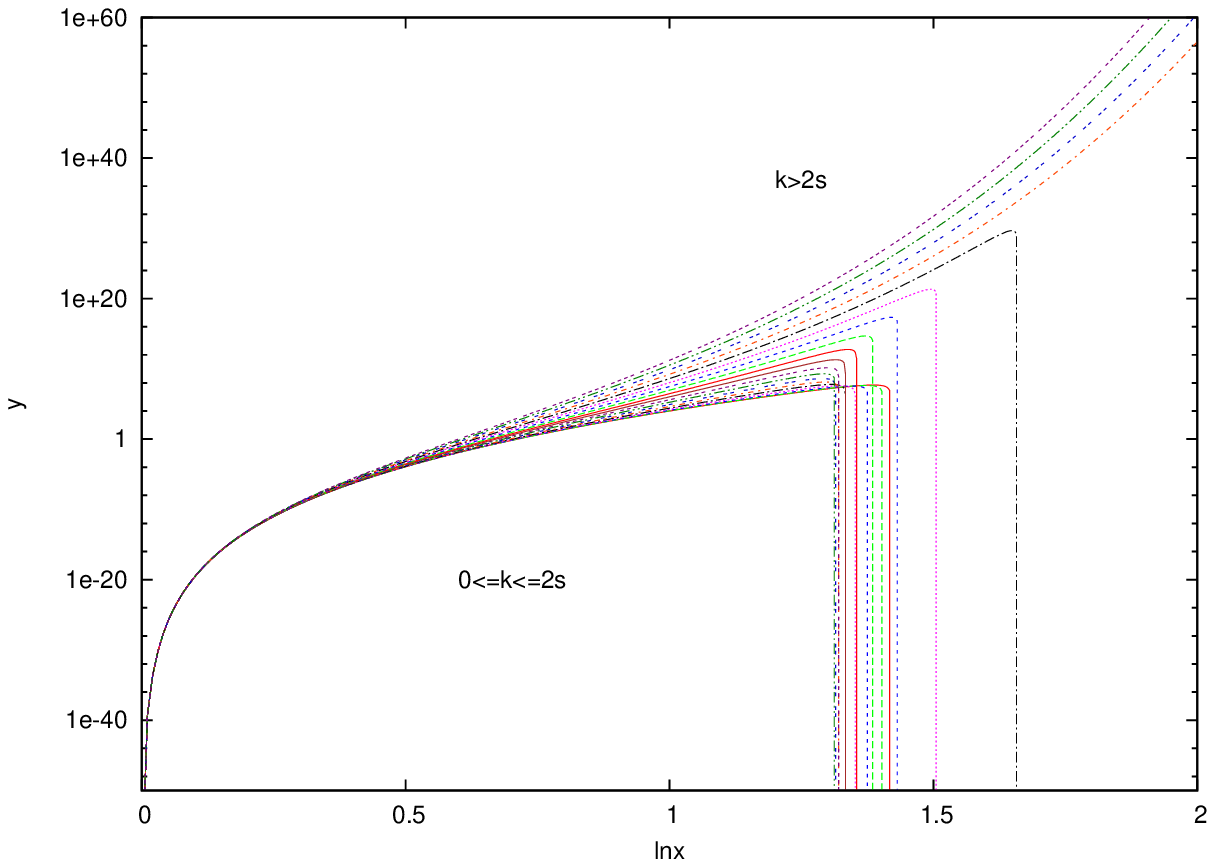}}
\hspace{2mm}
	\psfrag{k>2s}{}
	\psfrag{0<=k<=2s}{}
	\resizebox{8cm}{6cm}{\includegraphics{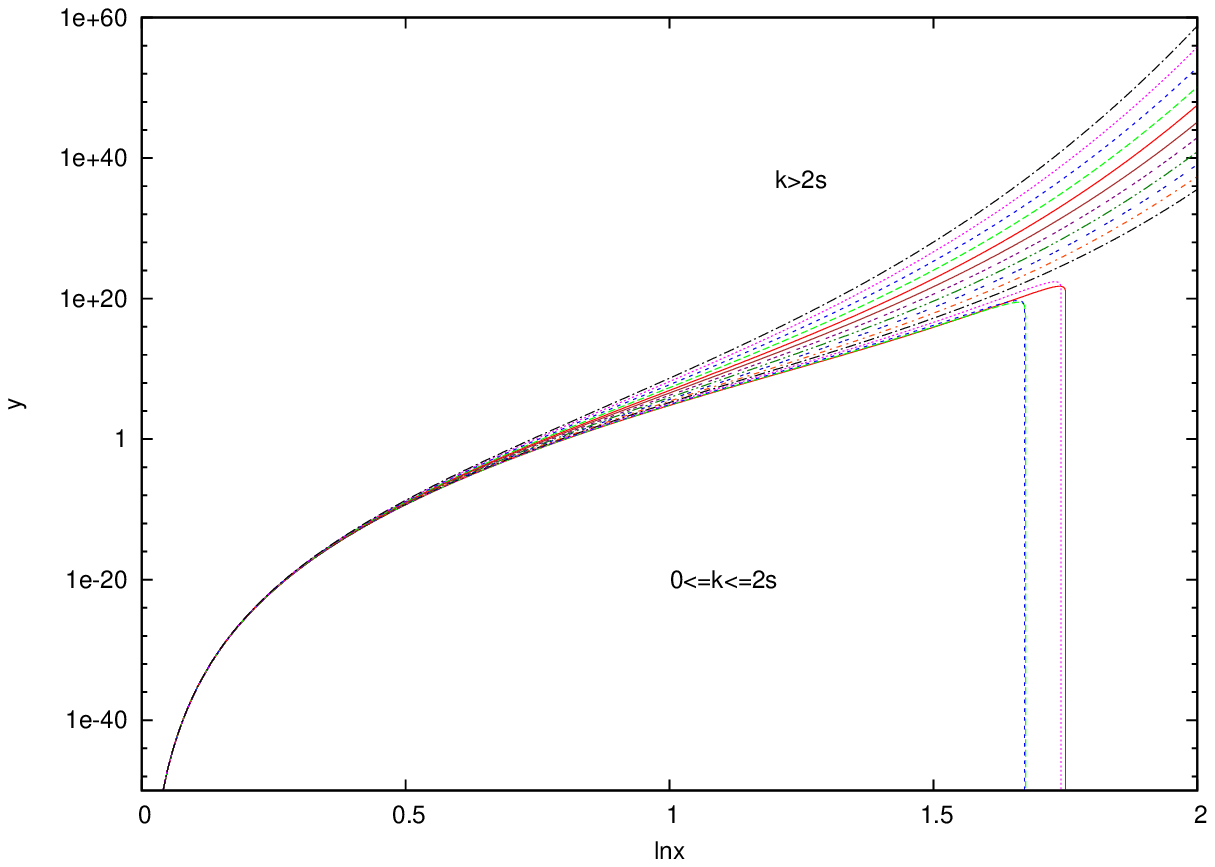}}

\hss}
\caption{The Jacobi determinant for  fluctuations 
around the $n=\nu=2$ vortex with ${\cal I}=0.87$ 
($\beta=2$, $\sin^2\thetaw=0.23$)  in the 
$m=0$ (left) and 
$m=2$ (right) sectors for different values of $\kappa$.  
}
\label{Fig_n=2}
\end{figure}

Calculating  $\Delta(\rho)$ is a much easier task than solving the boundary value problem
\eqref{SCHRODINGER}, since this simply  requires to  integrate the equations starting 
from $\rho=0$ with the boundary conditions  \eqref{BC_AXIS}. 
This should be done, in principle,  for each pair of values 
$\kappa,m$. In \cite{JGMV} this  method
was used to test stability  in the semilocal limit,
where $\thetaw=\pi/2$, while    
the typical behavior of the Jacobi determinant $\Delta(\rho)$ for 
$\thetaw< \pi/2$ 
is shown in 
 Figs.\ref{Fig_n=1},\ref{Fig_n=2}.

The main observation is as follows: 
the fundamental vortex with $n=\nu=1$ has one negative mode in the $m=0$
sector  for  every value of $\kappa$ from the interval
\be
|\kappa|<\kappa_{\rm max}({\cal I}). 
\ee 
This can be seen in Fig.\ref{Fig_n=1} where 
$\Delta(\rho)$ passes through
zero exactly once if $\kappa$ is small and never vanishes 
if $\kappa$ is large. In fact, 
the symmetry relations \eqref{TRANS_PARAM} 
imply that $\omega^2(-\kappa,-m)=\omega^2(\kappa,m)$,
so that
for $m=0$ one has $\omega^2(-\kappa)=\omega^2(\kappa)$ 
and it is therefore sufficient to consider 
only the $\kappa\geq 0$ region. 
In the Z string limit, for ${\cal I}=0$, one finds  
\be
\kappa_{\rm max}(0)=2\sigma(0)
\ee
(see Table I) 
and also  $\omega^2(0)=0$, so that the 
$\kappa=0$ mode is not negative. 
For ${\cal I}\neq 0$ one has 
\be
\kappa_{\rm max}({\cal I})> 2\sigma({\cal I}),
\ee
and in addition we find that the 
$\kappa=0$ mode is negative for $\thetaw\neq\pi/2$,
\be
\omega^2(0)<0,~~~~~{\cal I}\neq 0,~~
\ee
while in the semilocal limit one has $\omega^2(0)=0$ 
for all values of ${\cal I}$ \cite{JGMV}. 

It seems that for the $n=\nu=1$ vortex there are no other instabilities. 
We have checked for different values of $\kappa$ that for $m=1,2$ 
there are no negative modes (see Fig.\ref{Fig_n=1}),
while further increasing $m$ increases the centrifugal energy thus 
rendering the existence of bound states less probable.
As a result, it seems that the $n=\nu=1$
vortices are unstable only in the $m=0$ 
sector and are stable with respect to any other perturbations.

We also considered vortices with higher 
winding numbers $n$ and $\nu$ and found that the 
axially symmetric sector remains unstable for all solutions we examined 
(this was checked up to $n=3$). 
An example is shown in Fig.\ref{Fig_n=2} for $n=\nu=2$. 
This instability is qualitatively the same as for the $n=\nu=1$ vortex,
it exists for $0<|\kappa|<\kappa_{\rm max}({\cal I})$.  
However, solutions with $n>1$ have additional instabilities in sectors 
with $m>1$ which can be interpreted as splitting modes. For example,
the $n=2$ solutions are also unstable in the $m=2$ sector (see 
Fig.\ref{Fig_n=2}), which apparently corresponds 
to breaking of the $n=2$ vortex into two $n=1$ vortices. 
Such splitting instabilities are less interesting for us,
and in what follows we shall concentrate on the intrinsic 
instability of the fundamental $n=1$ vortex.

\subsection{Finding the eigenvalue}

Having detected the negative modes, 
we now wish to construct them explicitly. 
 Such a construction is considerably more involved 
than applying the Jacobi criterion, since it requires to solve  the boundary 
value problem for the $16$ coupled equations 
\eqref{SCHRODINGER} with the boundary conditions 
\eqref{BC_AXIS} and \eqref{GLOBAL}.
Unfortunately, even for $m=0$ these equations 
do not simplify much.
We solve them with
the multiple shooting method \cite{Stoer}, which requires to match at a fitting point 
the values of the 16 functions and their 16 first derivatives. 
It is then important to have enough free parameters in our disposal,
and in fact we have the $16$ integration constants in the local solutions \eqref{BC_AXIS},
then $16$ other constants in \eqref{GLOBAL}, and also 
the eigenvalue $\omega^2$.  As 
we consider a linear system, one constant can be fixed by the 
overall renormalization, so that there remain 32 parameters 
to fulfill the $32$ matchings conditions.
Resolving these conditions 
gives us the global solution $\Psi(\rho)$ of Eqs.\eqref{SCHRODINGER} in the interval
$\rho\in[0,\infty)$ and also the eigenvalue $\omega^2$.

\begin{figure}[ht]
\hbox to \linewidth{ \hss
	\psfrag{omega2}{$\omega^2$}
	\psfrag{k}{$\kappa$}
	\psfrag{s=0.7098}{${\cal I}=0$}
	\psfrag{s>sstar}{$\sigma>\sigma_\star$}
	\psfrag{s<sstar}{$\sigma<\sigma_\star$}
	
\psfrag{s=0.70}{${\cal I}=0.08$}
\psfrag{s=0.65}{${\cal I}=0.48$}
\psfrag{s=0.60}{${\cal I}=0.87$}
\psfrag{s=0.55}{${\cal I}=1.24$}

\psfrag{s=0.50}{${\cal I}=1.60$}
\psfrag{s=0.45}{${\cal I}=1.94$}
\psfrag{s=0.40}{${\cal I}=2.30$}

\psfrag{s=0.35}{${\cal I}=2.67$}
\psfrag{s=0.30}{${\cal I}=3.08$}
\psfrag{s=0.25}{${\cal I}=3.55$}
\psfrag{s=0.20}{${\cal I}=4.13$}
\psfrag{s=0.15}{${\cal I}=4.83$}

	\resizebox{8cm}{6cm}{\includegraphics{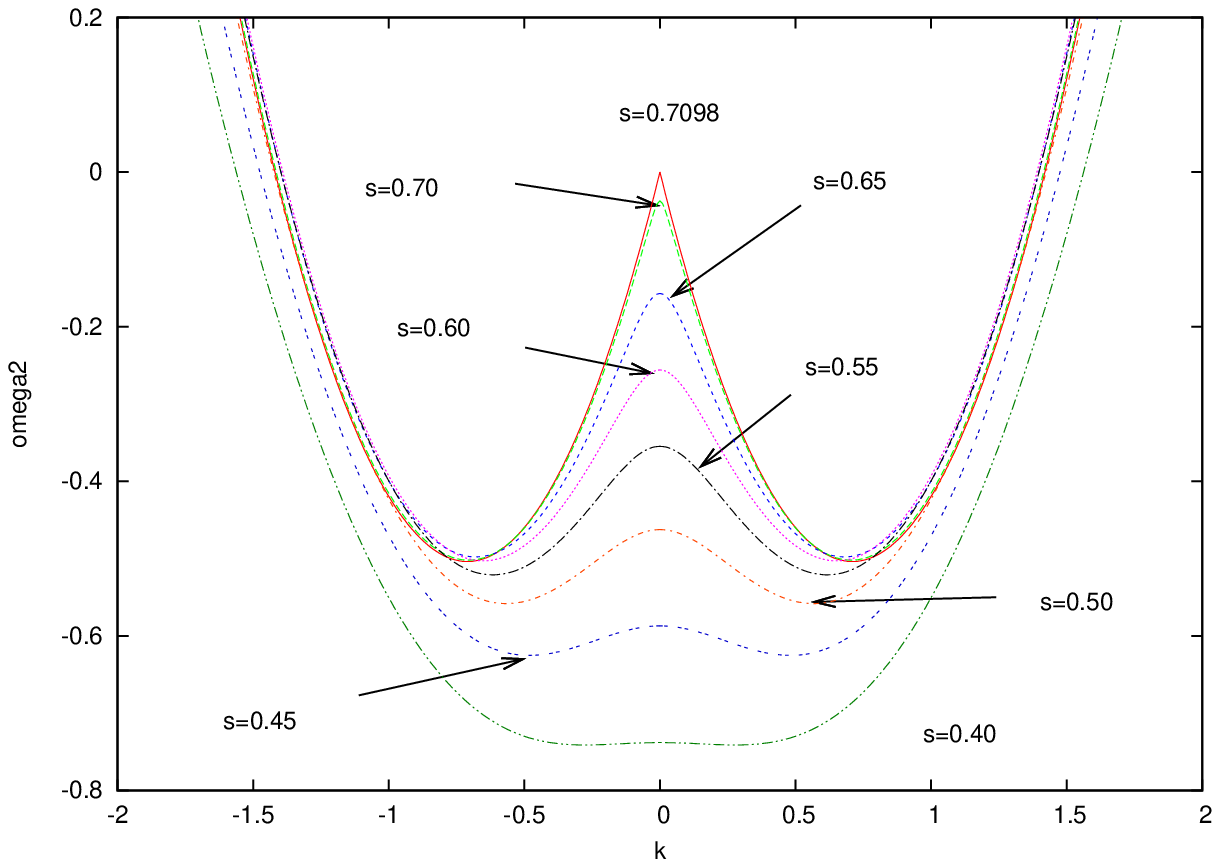}}
\hspace{2mm}
\resizebox{8cm}{6cm}{\includegraphics{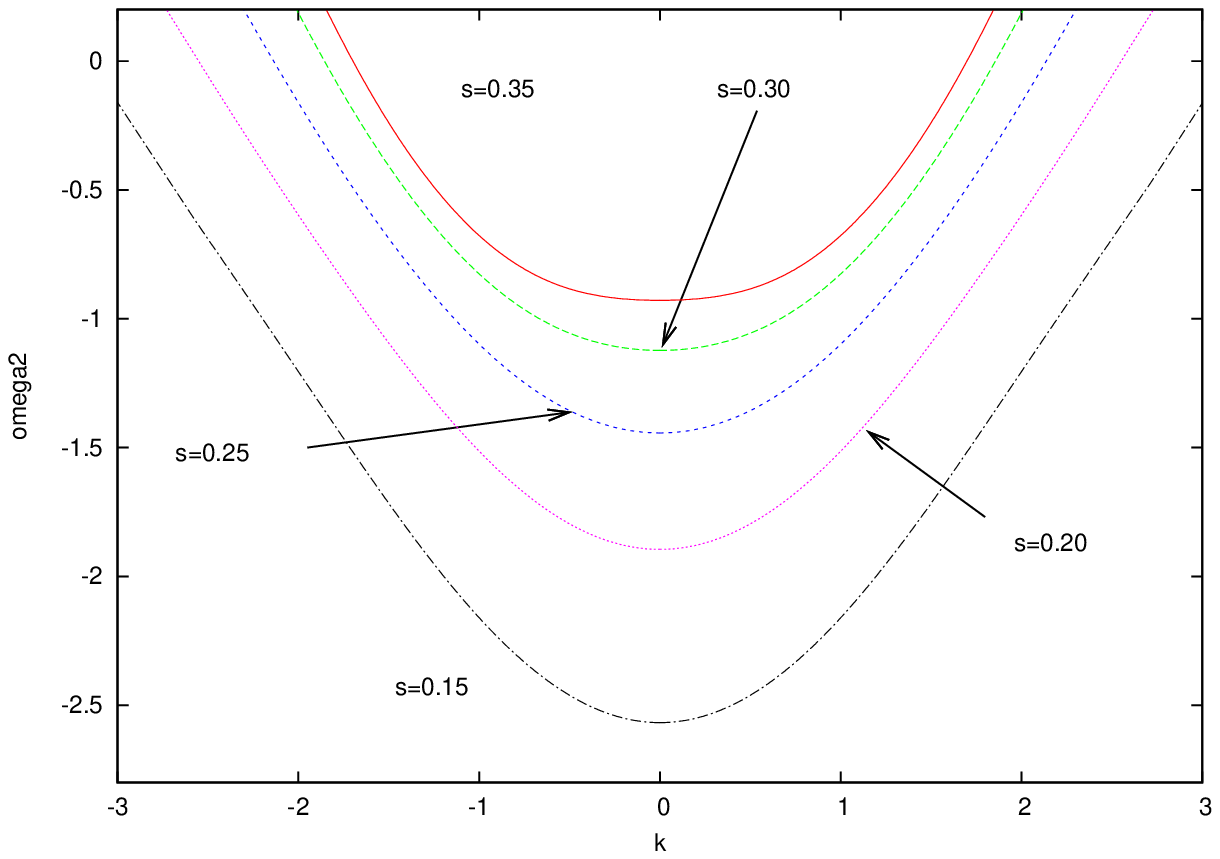}}

\hss}
\caption{\small Dispersion relation $\omega^2(\kappa)$ for the $m=0$ bound state 
solutions of Eqs.\eqref{SCHRODINGER} for the $n=\nu=1$ vortex 
($\beta=2$, $\sin^2\thetaw=0.23$) for 
${\cal I}<{\cal I}_\star=2.57$ (left) 
and for ${\cal I}>{\cal I}_\star$ (right).}
\label{FigDISP}
\end{figure}

As a result, we obtain the 
dispersion relation $\omega^2(\kappa)$  shown in 
Fig.\ref{FigDISP}. We see that 
there is a value $\kappa_{\rm max}({\cal I})$ such that 
$\omega^2(\kappa)<0$  for
$|\kappa|<\kappa_{\rm max}$.
For 
small currents the function $\omega^2(\kappa)$  has a double-well shape, with two minima 
of equal depth at $\kappa=\pm\kappa_{\rm min}$ 
and a local negative maximum at $\kappa=0$. As the current increases, 
$\kappa_{\rm min}$
decreases, the value $\omega^2(0)$
approaches  $\omega^2(\pm\kappa_{\rm min})$, and finally 
$\kappa_{\rm min}$ vanishes
for ${\cal I}={\cal I}_\star$ 
when  all three extrema 
of $\omega^2(\kappa)$ merge into a global minimum. 
For ${\cal I}>{\cal I}_\star$  
the function $\omega^2(\kappa)$ shows only one global minimum at $\kappa=0$. 
Some numerical characteristics of 
$\omega^2(\kappa)$ are presented in Table \ref{TABLE}.

\begin{figure}[ht]
\hbox to \linewidth{ \hss
	\psfrag{kmin}{\large$\kappa_{min}$}
	\psfrag{I}{${\large\cal I}$}
	\psfrag{Istar}{\large${\cal I}_\star$}
	\resizebox{8cm}{6cm}{
\includegraphics{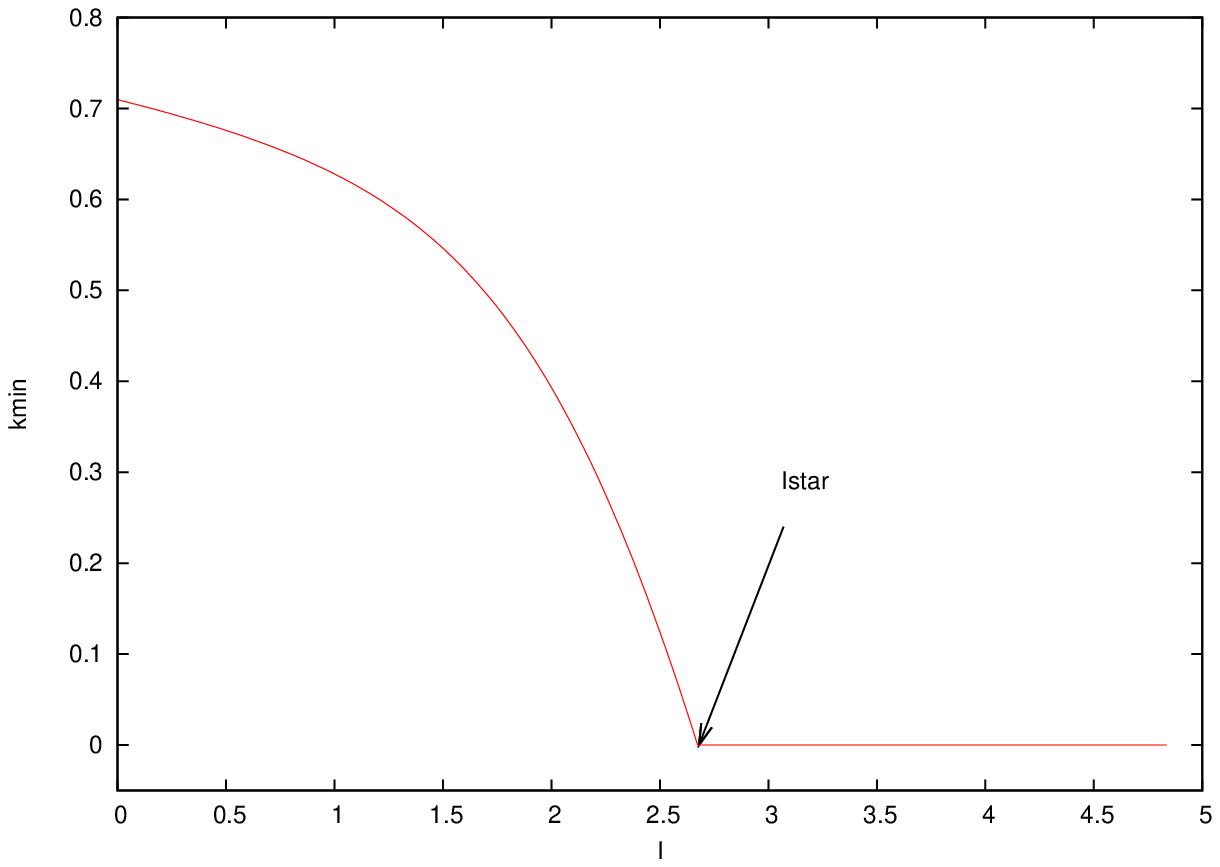}}
\hspace{2mm}
	\psfrag{sigma}{\large$\sigma$}
	\psfrag{q}{\large$q$}
	\psfrag{Zstring}{\large${\cal I}=0$}
	\psfrag{sstar}{\large ${~~~\cal I}_\star$~~}
	\psfrag{I_inf}{\large${\cal I}\to\infty$}
	\psfrag{transition}{}
	\resizebox{8cm}{6cm}{\includegraphics{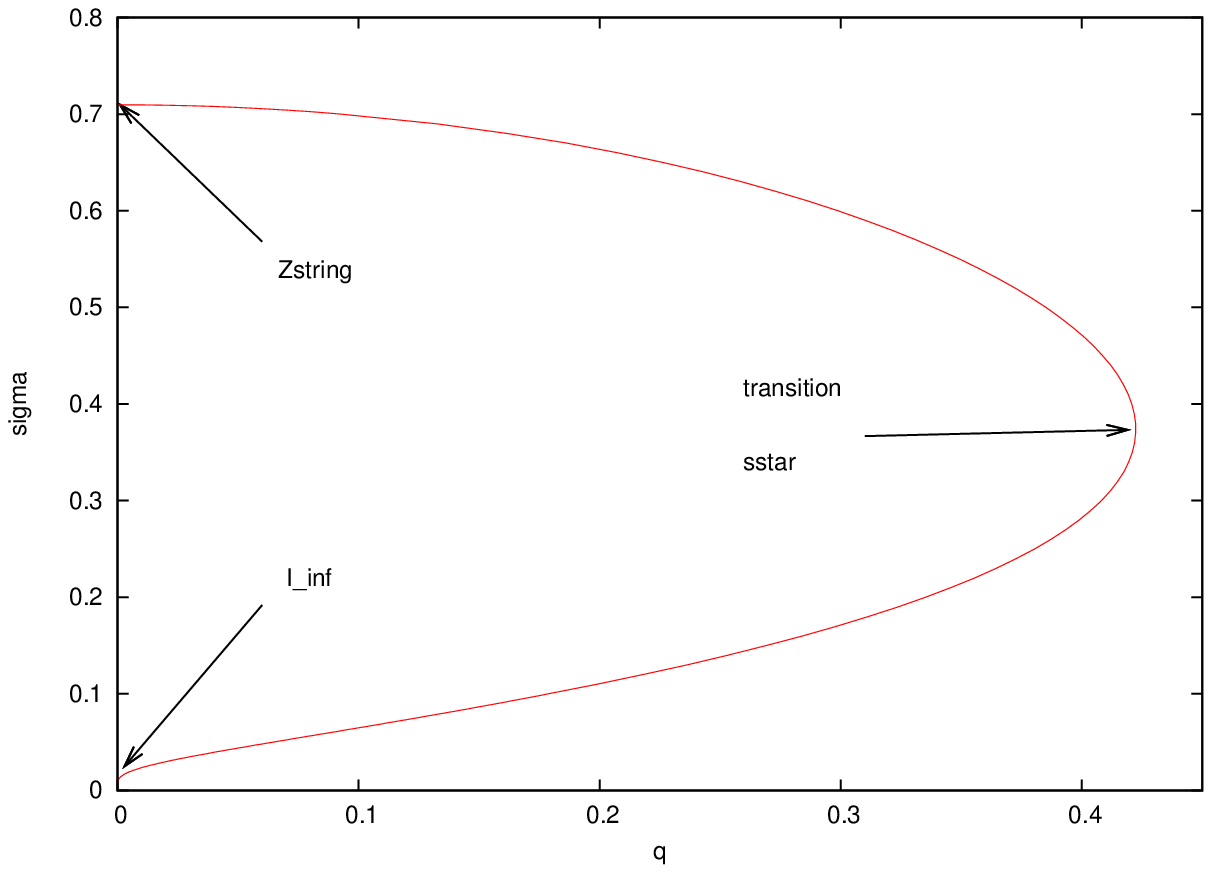}}
\hss}
\caption{Profiles of $\kappa_{\rm max}({\cal I})$ (left) and $\sigma({\cal I})$ against 
$q({\cal I})$ (right) for the same vortex solution as in Fig.\ref{FigDISP}. 
}
\label{Fig_Disp_param}
\end{figure}

The passage from the two-well to one-well structure of the dispersion relation 
suggests that the system undergoes some kind of phase transition at
${\cal I}={\cal I}_\star$. This is corroborated by the profile 
of $\kappa_{\rm min}({\cal I})$ (see Fig.\ref{Fig_Disp_param})
reminding of a second order phase transition. The point ${\cal I}={\cal I}_\star$ 
is also distinguished by the fact that the background `consensate parameter' 
$q=f_2(0)$ attains its maximal value there  (see Fig.\ref{Fig_Disp_param}). 
When ${\cal I}$ grows further, $q$ starts decreasing and tends to zero as ${\cal I}\to\infty$. 
For large currents the vortex shows in its central part an unbroken phase region where the Higgs field
is driven to zero by the strong magnetic field \cite{JGMV2}. 
This suggest that the point ${\cal I}={\cal I}_\star$  
corresponds to the transition in which the unbroken phase just starts to appear in the vortex center.  
The plot $\sigma(q)$ shows a characteristic two-branch structure (see Fig.\ref{Fig_Disp_param}) and
the point ${\cal I}={\cal I}_\star$ corresponds to the bifurcation between the two branches. 
Although this suggest that the stability may change at this point, we know already  that the number of 
instabilities remains actually the same, but  the 
dispersion relation changes its shape. As discussed in  Sec.\ref{cvort} below, this should 
alter  the 
generic instability pattern.

\begin{table}[!ht]
\caption{Parameter values 
for the $n=\nu=1$ vortices with $\beta=2$, $\sin^2\thetaw=0.23$. }
\begin{center}\begin{tabular}{|c|c|c|c|c|c|}
\hline
${\cal I}$ & $\sigma$ & $\omega^2(0)$ & 
$\kappa_{\rm{min}}$ & $\omega^2(\kappa_{\rm min}) $ & $\kappa_{\rm max}$ \\
\hline\hline
0 & 0.709697	&0.0    	&0.709697	&-0.503670	&1.419394 \\
0.0804 & 0.700   	&-0.0370976	&0.705  	&-0.519740	&1.415    \\
0.4851 & 0.650   	&-0.157024	&0.680  	&-0.497821	&1.395    \\
0.8739 & 0.600   	&-0.255942	&0.655  	&-0.502902	&1.390    \\
1.2430 & 0.550   	&-0.354806	&0.615  	&-0.520995	&1.400    \\
1.6002 & 0.500   	&-0.462475	&0.570  	&-0.558202	&1.425    \\
1.9494 & 0.450   	&-0.587058	&0.475  	&-0.625174	&1.475    \\
2.3004 & 0.400   	&-0.738065	&0.280  	&-0.741152	&1.560    \\
2.6740 & 0.350   	&-0.928761	&0.0    	&-0.928761	&1.695    \\
3.0831 & 0.300   	&-1.12311	&0.0    	&-1.12311.	&1.855    \\
3.5594 & 0.250   	&-1.44332	&0.0    	&-1.44332.	&2.135    \\
4.1327 & 0.200   	&-1.89531	&0.0    	&-1.89531.	&2.550    \\
4.8335 &0.150   	&-2.56766	&0.0    	&-2.56766.	&3.150     \\
\hline
\end{tabular}\end{center}
\label{TABLE}
\end{table}

Let us now consider the limiting cases where the vortex current 
is either  small or large. 
This will help to understand the structure of  curves  in Fig.\ref{FigDISP}.

\section{Zero current limit  \label{APP_C}}

When the vortex current tends to zero, the solutions reduce to 
Z strings \cite{Zstring}, whose stability has been studied before  \cite{Goodband}, \cite{James}. 
The most detailed consideration of the problem was presented in Ref.\cite{Goodband},
whose results we have been able to confirm. 

In zero current limit the vortex field amplitudes become 
\begin{align}
\Y&=-1,~~~~\Z=2g^{\prime 2}(v_{\mbox{\tiny ANO}}-\n)+2\n-\nu\equiv  v_{\mbox{\tiny Z}},~~~~
\Om_1=0,~~~~\Om_3=1, \nonumber \\
\W_1&=0,~~~~~~
\W_3=2g^{2}(v_{\mbox{\tiny ANO}}-\n)+\nu\equiv v_{\mbox{\tiny Z}3},~~~~
\f=f_{\mbox{\tiny ANO}}\equiv  f_{\mbox{\tiny Z}},~~~~\p=0 ,              \label{Zsol}
\end{align}
and the field
equations \eqref{ee1}--\eqref{CONS1} reduce to 
 the ANO system 
\begin{align}                                           
\frac{1}{\rho}(\rho f_{\mbox{\tiny ANO}}^\prime)^\prime&=
\left(
\frac{v^2_{\mbox{\tiny ANO}}}{\rho^2}
+\frac{\beta}{4}(f_{\mbox{\tiny ANO}}^2-1)
\right) f_{\mbox{\tiny ANO}}\,,                       \notag \\
 \rho\left(\frac{v_{\mbox{\tiny ANO}}^\prime}{\rho}\right)^\prime&
=\frac{1}{2}\,
f_{\mbox{\tiny ANO}}^2\,v_{\mbox{\tiny ANO}}               \label{ANOeqs} 
\end{align}
whose solutions fulfill the boundary conditions $0\leftarrow f_{\mbox{\tiny ANO}}\to 1$ and 
$n\leftarrow v_{\mbox{\tiny ANO}}\to 0$ as $0\leftarrow \rho\to \infty$. 
The solutions depend  only on the winding number $n$, although 
when written in the gauge \eqref{003} the fields also contain $\sigma_\alpha,\nu$,
\be                           \label{003Z}
{\cal W}_{Z}=(\tau^3-1)\,\om_\alpha dx^\alpha -
[v_{\mbox{\tiny Z}}(\rho)+               
 \tau^3 v_{\mbox{\tiny Z}3}(\rho)]\, d\varphi,
~~~~~~~~
\Phi_{Z}=\left(\begin{array}{c}
f_{\mbox{\tiny Z}}(\rho) \\
0
\end{array}\right).       
\ee
The values of $\sigma^2,\nu$ are determined by those for the generic vortices
in the ${\cal I}\to 0$ limit, one has for example $\sigma^2=\sigma^2(\beta,\thetaw,n,\nu)>0$. 
Although 
$\sigma_\alpha,\nu$ can be gauged away for this solution, they reappear 
again in the perturbation equations. 
In particular, $-\sigma^2$ determines (see \eqref{lamlam}) 
the eigenvalue in Eqs.\eqref{SOo}, and 
since it is is negative,  Z strings 
are unstable. Stable Z strings also exist, for unphysical values of 
$\beta,\thetaw$ (the eigenvalue is then positive), but they cannot be viewed as
limits of superconducting vortices \cite{JGMV2}, 
so that they are not relevant for us. 
One can accurately determine the parameter regions in the 
$\beta,\thetaw$ plane where Z strings are unstable/stable and so can/cannot 
be promoted to the superconducting vortices by studying solutions of 
Eqs.\eqref{SOo} with $\sigma^2=0$
\cite{Goodband}, \cite{JGMV2}.

Imposing \eqref{Zsol}, 
the potential energy matrix in the Schr\"odinger operator \eqref{SO} 
becomes block diagonal, so that the space of perturbations spanned by the 
16-component vector $\Psi$ in \eqref{PSI_U} decomposes into a direct sum 
  of
six  one-dimensional subspaces, 
one four-dimensional subspace, and two three-dimensional 
subspaces. 
The six one-dimensional subspaces are spanned by 
${\cal A}_{\pm 1}$, ${\cal A}_{0}$, ${\cal Z}_{0}$, ${\cal W}_{0}^{\pm}$,
which describe the photon and the longitudinal components of  Z and W bosons. 
The potentials in the corresponding one-channel  Schr\"odinger equations are 
positive definite so that there are no negative modes in these sectors.

The four-dimensional subspace 
is spanned by ${\cal Z}_{\pm},h_1^{\pm}$, which correspond to
the transverse components of  Z and Higgs bosons. For $m=0$ this 
space further splits into  sectors spanned, respectively, 
by ${\cal Z}_{+}+{\cal Z}_{-}$, $h_1^{+}+h_1^{-}$ and by 
${\cal Z}_{+}-{\cal Z}_{-}$, $h_1^{+}-h_1^{-}$. Both of them contain 
bound states with $\omega^2>0$ (in the first sector they exist only for $\beta<1.5$)
but there are no negative modes in this case. 

\begin{figure}[ht]
\hbox to \linewidth{ \hss

	\psfrag{x}{$\kappa$}
	\psfrag{b}{$k_{+}=-\sigma$}
	\psfrag{b1}{$k_{+}=\sigma$}
	\psfrag{b2}{$k_{-}=\sigma$}
	\psfrag{b3}{$k_{-}=-\sigma$}
	\psfrag{o}{$\omega^2=(\kappa\pm\sigma)^2-\sigma^2$}

	\resizebox{8cm}{5cm}{\includegraphics{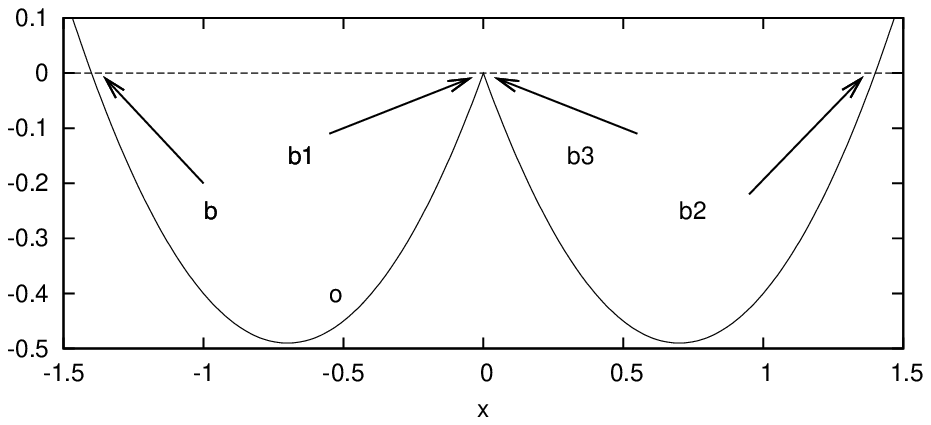}}

\hspace{2mm}

	\psfrag{x}{$k$}
	\psfrag{y}{$\omega^2$}
	\psfrag{b}{$k=\pm\sigma$}
	\psfrag{b1}{bifurcations}
	\psfrag{o}{$\omega^2=k^2-\sigma^2$}
	\resizebox{6cm}{5cm}{\includegraphics{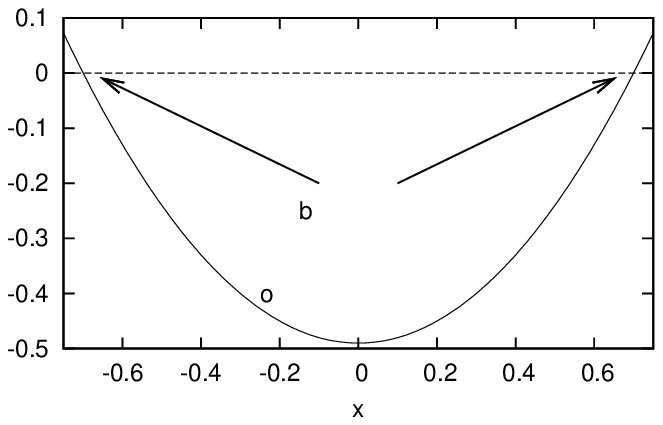}}

\hss}
\caption{\small Dispersion relation \eqref{omk} for the bound state solutions of the 
eigenvalue problem \eqref{SOo} (left). The two parabolas correspond to solutions in the 
independent
$\Psi_{+}$
and $\Psi_{-}$ 
 sectors.   Passing to the gauge \eqref{003aZ}
they get mapped into one parabola giving the dispersion relation for the modes \eqref{i1}
(right). The arrows indicate bifurcations with the superconducting branch. 
}
\label{Fig_Zdisp}
\end{figure}

The remaining three-dimensional subspaces
spanned by  ${\cal W}^\pm_{+}, h_2^{+}$ 
and  ${\cal W}^\pm_{-},h_2^{-}$  contain the negative modes. 
The perturbations are governed in this case by 
\be                  \label{SOo}
-\frac{1}{\rho}\left(\rho\Psi_\pm^\prime\right)^\prime+\mathcal{U}_\pm\Psi_\pm=\Lambda_\pm\Psi_\pm
\ee
with 
$
\Lambda_\pm=\omega^2-(\sigma\mp\kappa)^2
$
 and 
\begin{align}
\Psi_\pm&=\left( \begin{array}{c} 
		\Wpmp\\
		\Wpmm\\
		\Htpm\\
	\end{array}\right) \, ,&
\mathcal{U}_\pm &=\left( \begin{array}{ccc} 
\Dwpmp	&0	&V^\pm	\\	
0	&\Dwpmm	&V^\mp	\\	
V^\pm	&V^\mp	&\Dhtpm	\end{array}\right) \, ,
\end{align}
where 
\begin{align}
\Dwpme &=\frac{\left(2g^2(\vano-n)
+\nu\pm(m-\nu)\right)^2}{\rho^2}\pm4\eta 
g^2\frac{\vano^\prime}{\rho}+\frac{g^2}{2}\fano	\, , \notag \\
\Dhtpm &=\frac{(\vano\mp m)^2}{\rho^2}+\frac{\beta}{4}\left(\fano^2-1\right)+\frac{g^2}{2}\fano	\, , ~~~~
V^\pm  =g\left(\fano^\prime\pm\frac{\vano\fano}{\rho}\right)	\, . 
\end{align}
For $m=0$  
equations \eqref{SOo}  admit bound state solutions both in the $\Psi_{+}$ 
and $\Psi_{-}$ subspaces with the eigenvalue
\be					\label{lamlam}
\Lambda_{+}=\Lambda_{-}=-\sigma^2\equiv-\sigma^2(\beta,\thetaw,n,\nu)<0
\ee 
 for $\nu=1,\ldots  \nu_{\rm max}$ 
where $n \leq \nu_{\rm max}(\beta,\thetaw,n)\leq 2n-1$ 
 \cite{JGMV2}.  
These
bound states  are characterized, respectively, by the 
dispersion relation 
\be                                          \label{omk}
\omega^2= \omega_{\pm}^2(\kappa)\equiv (\sigma\mp\kappa)^2-\sigma^2=
\kappa(\kappa\mp2\sigma). 
\ee 
One has $\omega_{+}^2(\kappa)<0$ for $0<\kappa<2\sigma$
and $\omega_{-}^2(\kappa)<0$ for $-2\sigma<\kappa<0$ so that 
there is one negative mode for every value of $\kappa$
from the interval 
$
(-2\sigma,0)\cup (0,2\sigma).
$
As a result, the dispersion relation for  negative modes 
is described by $\omega_{+}^2(\kappa)$
for $\kappa>0$ and by $\omega_{-}^2(\kappa)$
for $\kappa<0$, therefore 
the $\omega^2(\kappa)$ curve 
consists of two parabolas intersecting at $\kappa=0$
(see Fig.\ref{Fig_Zdisp}). These parabolas continue to the  
$|\kappa|>2\sigma$ regions where there are bound states with $\omega^2>0$.  
However, they should terminate for $\kappa=0$, since the effective photon mass 
 $\mu^2_{\mbox{\tiny $A$}}=\kappa^2-\omega^2=\pm2\sigma\kappa$ defined by Eq.\eqref{MASSES} 
becomes imaginary after this point. Although the photon 
 decouples for exactly vanishing background current, it rests coupled 
for however small but non-zero currents, when the background is arbitrarily close
to Z string.

Let us now use \eqref{DECOMP} to reconstruct the dependence of negative modes 
on all spacetime coordinates. Then we apply to \eqref{003Z} the gauge transformation 
${\rm U}=e^{in\varphi}{\rm u}(\nu\varphi){\rm u}(\sigma_\alpha x^\alpha)$ with 
${\rm u}(X)\equiv e^{iX(1-\tau^3)/{2}}$.
The Z string   becomes then globally regular and independent of $\sigma_\alpha,\nu$, 
\be                           \label{003aZ}
{\cal W}_Z^{\rm reg}=
2(g^{\prime 2}+g^2\tau^3)(\n-v_{\mbox{\tiny ANO}}(\rho))\,d\varphi,      
~~~~ 
\Phi_Z^{\rm reg}=
\left(
\begin{array}{c}
e^{in\varphi}f_{\mbox{\tiny ANO}}(\rho)\\
0 
\end{array}  
\right),     
\ee
in which form it is usually described in the literature \cite{Zstring}. 
The negative modes read in this gauge   
(writing down only the Higgs field perturbations) 
%\be                                                     \label{i1}
$
\delta\Phi_2=C_{+}h_2^{+}(\rho)e^{|\omega_{+}(k_{+})|t}e^{-ik_{+}z}
$
for $\kappa\in(0,2\sigma)$ and 
$                                                    
\delta\Phi_2=C_{-}h_2^{-}(\rho)e^{|\omega_{-}(k_{-})|t}e^{ik_{-}z}
$
for $\kappa\in(-2\sigma,0)$. 
Here $C_{\pm}$ are integration constants, $k_{\pm}=\kappa\mp \sigma$ 
and $\omega^2_{\pm}={k_{\pm}^2-\sigma^2}$. 
Replacing  $k_{\pm}\to k$ and using the fact that $h_2^{+}(\rho)=h_2^{-}(\rho)$
one can write these solutions simply as  
\be                                                     \label{i1}
\delta\Phi_2=C_{\pm}h_2^{+}(\rho)e^{|\omega(k)|t}e^{\mp ikz}
\ee
with $\omega^2={k^2-\sigma^2}$ (see Fig.\ref{Fig_Zdisp}).
 These negative modes can be viewed as standing waves of length
$\lambda=2\pi/k$ whose amplitude grows in time. 
For $k=\pm\sigma$ one obtains zero modes corresponding to the bifurcations
of Z strings 
with the superconducting solutions. 
Since the minimal wavelength of negative modes is 
$\lambda_{\rm min}=2\pi/\sigma$, this suggests that the instability could be removed by 
imposing  periodic boundary conditions along the $z$-axis 
with the period $L\leq \lambda_{\rm min}$. 
However, this would not remove the {homogeneous} $k=0$  mode, 
$\delta\Phi_2=Ch^{+}_2(\rho)e^{\sigma t}$,
since it is independent of $z$ and so can be considered as periodic with any period. 

Let us, however, consider Z string in yet another gauge -- the one given by Eq.\eqref{003a}.
In this gauge the fields are also globally regular 
(we assume the restframe condition $\sigma_\alpha=\sigma\delta^3_\alpha$), 
\be                                    \label{003aZ1}
{\cal W}=\sigma(\tau^3-1)dz+{\cal W}_Z^{\rm reg},~~~~~~~~~~~ 
\Phi=\Phi_{Z}^{\rm reg}\,.
\ee
The correspondence between this gauge and \eqref{003aZ} is provided by 
 the gauge transformation with 
\be                        \label{UUU}
{\rm U}={\rm u}(\sigma z)=\left[\begin{array}{cc}
1 & 0 \\
0 & e^{i\sigma z}   
\end{array}\right].
\ee
The negative modes \eqref{i1} now become
\be                                                     \label{i4}
\delta\Phi_2=C_{\pm}h_2^{+}(\rho)e^{|\omega_{\pm}(\kappa)|t}e^{\mp i\kappa z}
\ee
with $\kappa\in(-2\sigma,0)\cup (0,2\sigma)$, 
these are standing waves of length 
$\lambda=2\pi/{\kappa}\geq\lambda_{\rm min}=\pi/\sigma$. 
Imposing now periodic boundary conditions with period 
$
L={\pi}/{\sigma}
$
will remove {\it all} negative modes. In particular, the mode which used to be 
homogeneous becomes now $z$-dependent with $\kappa=\pm\sigma$, so that it 
will be removed.  The $z$-independent mode
now corresponds to $\kappa=0$ and it will not be removed, 
but this mode is {\it not negative},
so it is harmless. 
One should say that in the case under consideration all gauge invariant quanitites 
like $\delta B_{\mu\nu}$ and $\delta(n^a{\rm W}^a_{\mu\nu})$ vanish and 
there is no gauge invariant way to decide which modes are homogeneous.

We notice finally 
that the gauge transformation \eqref{UUU} is {\it not} periodic in the interval
$[0,\pi/\sigma]$, and therefore imposing the periodicity breaks the gauge equivalence between \eqref{003aZ}
and \eqref{003aZ1}. The two descriptions of Z string become therefore physically different, 
which is why \eqref{003aZ1} becomes stable upon imposing the periodicity
 while \eqref{003aZ} rests unstable. To the best of our knowledge, such a possibility 
to stabilize Z strings has never been discussed in the literature.

Since the Z string zero modes for $\kappa=\pm 2\sigma$
correspond to bifurcations 
with the superconducting solutions, they  
can be viewed as small deformations induced by the current.  
Now, one has $\omega^2(\pm\kappa_{\rm max})=0$ also for ${\cal I}\neq 0$, which suggests 
that the related zero modes also correspond to deformations 
induced by a small current variation. However, for ${\cal I}\neq 0$ such deformations  
would inevitably contain 
logarithmically growing at infinity terms and therefore would not correspond to bound state
solutions of the perturbation equations.  This suggests that the $\kappa=\pm \kappa_{\rm max}$
zero modes could
correspond to variations with respect to some other parameter. In other words,
it may be that the vortex solutions admit stationary generalizations within 
a field ansatz more general than  \eqref{003}.

\section{Large current limit \label{S_LARGE}}

When the current ${\cal I}$ is large, the vortex develops in its center a region of size 
$\sim{\cal I}$ where the magnetic field is so strong that it quenches the Higgs field 
to zero. Most of this region  is filled with 
the massless electromagnetic and Z fields produced by the current. The latter is 
carried by the  charged W boson condensate 
confined in the compact  core of size $\sim 1/{\cal I}$ placed 
in the very center of the symmetric phase. 
Outside the symmetric 
phase region the Higgs field relaxes to its vacuum value and everything reduces to the 
ordinary electromagnetic Biot-Savart  field \cite{JGMV2}.

The vortex fields in this limit can  be 
described by splitting the space into two 
parts: the core region $\rho<{x_0}/{\cal I}$
and the exterior region $\rho>{x_0}/{\cal I}$. 
The  fields in the core can be approximated by 
\begin{align}\label{CORE}
&f_1=f_2=\sigma u_3=v_1=0\, ,~~ 
\sigma u=const.	\, ,~~  
v =1	\, ,  \notag \\
&\sigma u_1(\rho)={\cal I} U_1({\cal I}\rho)\, ,~~~~~~~~~ 
v_3=V_3({\cal I}\rho)	\, ,
\end{align}
in which case the field equations \eqref{ee1}--\eqref{CONS1} reduce to 
\begin{align}                                \label{uv:0}                                
\frac{1}{x}(x U_1^\prime)^\prime&=\frac{V_3^2}{x^2}\,U_1, ~~~~~~~~~~~~~~~
{x}\left(\frac{V_3^\prime}{x}\right)^\prime=U_1^2 V_3,\
\end{align}
with $x={\cal I}\rho$.
The solution of these equations exhibits the following behavior
for $0\leftarrow x\to \infty$,
\be
0\leftarrow U_1(x)\to a\ln x+b,~~~~~~~~ 
1\leftarrow V_3(x)\to 0
\ee 
(here $a=0.29,b=-0.08$ if $g^2=0.23$) where the large $x$ asymptotic 
is attained, up to exponentially small terms,  
at $x\equiv x_0\approx 10$. This determines the size of the core region. 
This solution describes the current-carrying 
charged W condensate confined in the core, 
the current value  
entering \eqref{CORE} as the scale
parameter.  

The fields for $\rho>{x_0}/{\cal I}$  
can be found separately and then matched to the core fields  
at $\rho={x_0}/{\cal I}$ \cite{JGMV2}. 
We do not need here the precise form of the 
$\rho>{x_0}/{\cal I}$ solutions,
since it is sufficient to analyze the stability of the 
core region. Indeed, suppose that we find 
a negative mode localized in the core. Since 
it vanishes in the outside region,
it fulfills the perturbation equations also there,
so that it will  be a negative mode of the whole 
vortex configuration. 
In principle there could be additional negative modes 
in the outside region, however, 
it turns out that  the core negative modes fit in well with the 
general instability pattern described above, which suggests that 
all vortex instabilities are localized in its core.   

To study the core instabilities, we inject \eqref{CORE}
into the perturbation equations \eqref{SO}. Passing to the radial 
variable  $x={\cal I}\rho$ and defining
\be
\tilde{\omega}=\omega/{\cal I},~~~~~~\tilde{\kappa}=\kappa/{\cal I}, 
\ee 
the current ${\cal I}$ 
drops from the equations. For $m=0$ the  equations  
split into three independent multichannel sectors plus free wave equations,
and applying the Jacobi criterion one can check that the negative modes
are contained only in the sector spanned by five amplitudes 
\begin{align}\label{LINEAR_COMB_LARGE}
Y^2_4 &\equiv {X_1(x)} \, , &
Y^3_2 &\equiv {X_2(x)} \, , &
X^3_4 &\equiv {X_5(x)} \, , \notag\\
\sqrt{2}\,X^1_3 &\equiv x\left(X_3(x)-X_4(x)\right) \, , &
\sqrt{2}\,X^2_2 &\equiv X_3(x)+X_4(x) \, .
\end{align}

\begin{figure}[ht]
\hbox to \linewidth{ \hss
	\psfrag{y}{}
	\psfrag{lnx}{$\ln(1+x)$}
	\resizebox{8cm}{6cm}{\includegraphics{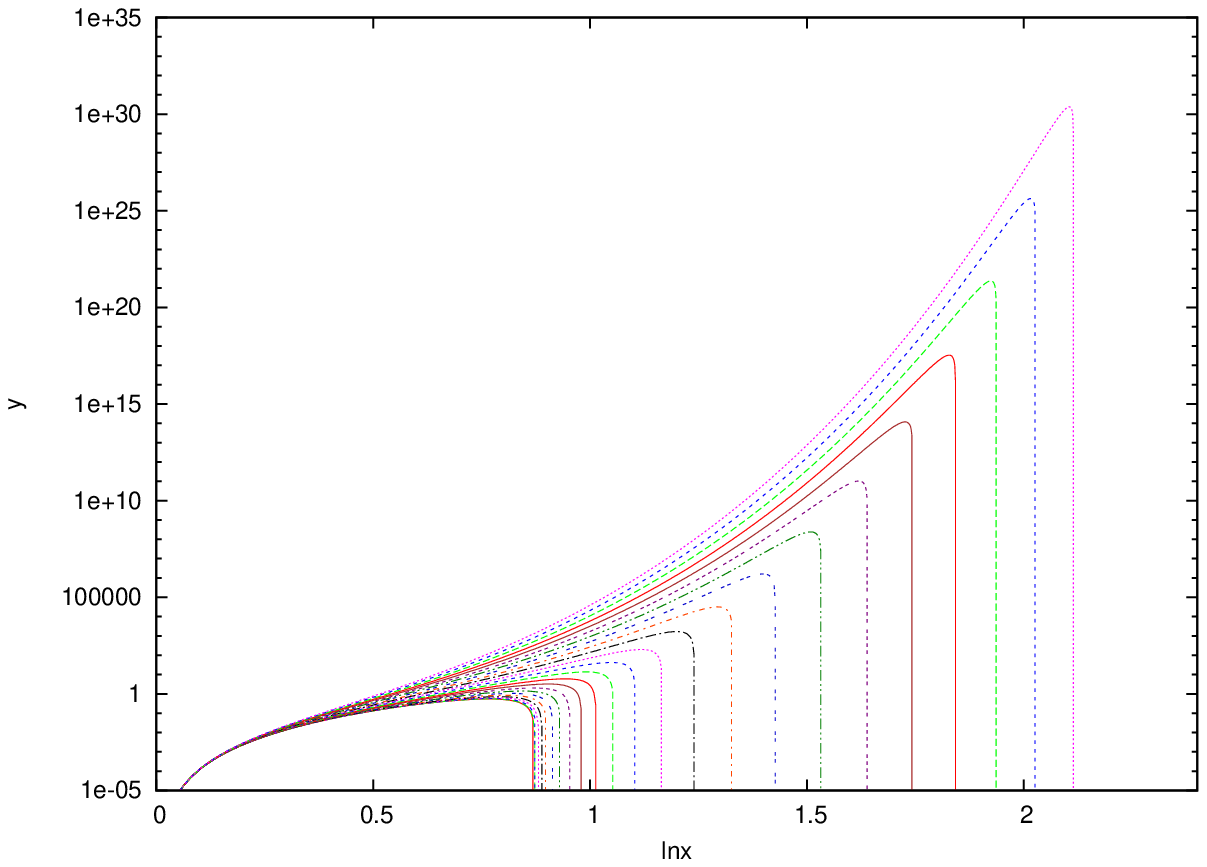}}

\hspace{2mm}
	\psfrag{omega2}{\large $\tilde{\omega}^2$}
	\psfrag{k2-omega2}{}
	\psfrag{k**2}{$K^2$}
	\psfrag{k}{\large $\tilde{\kappa}$}
	\psfrag{k**2-omega**2}{$K^2-\Omega^2$}
	\resizebox{8cm}{6cm}{\includegraphics{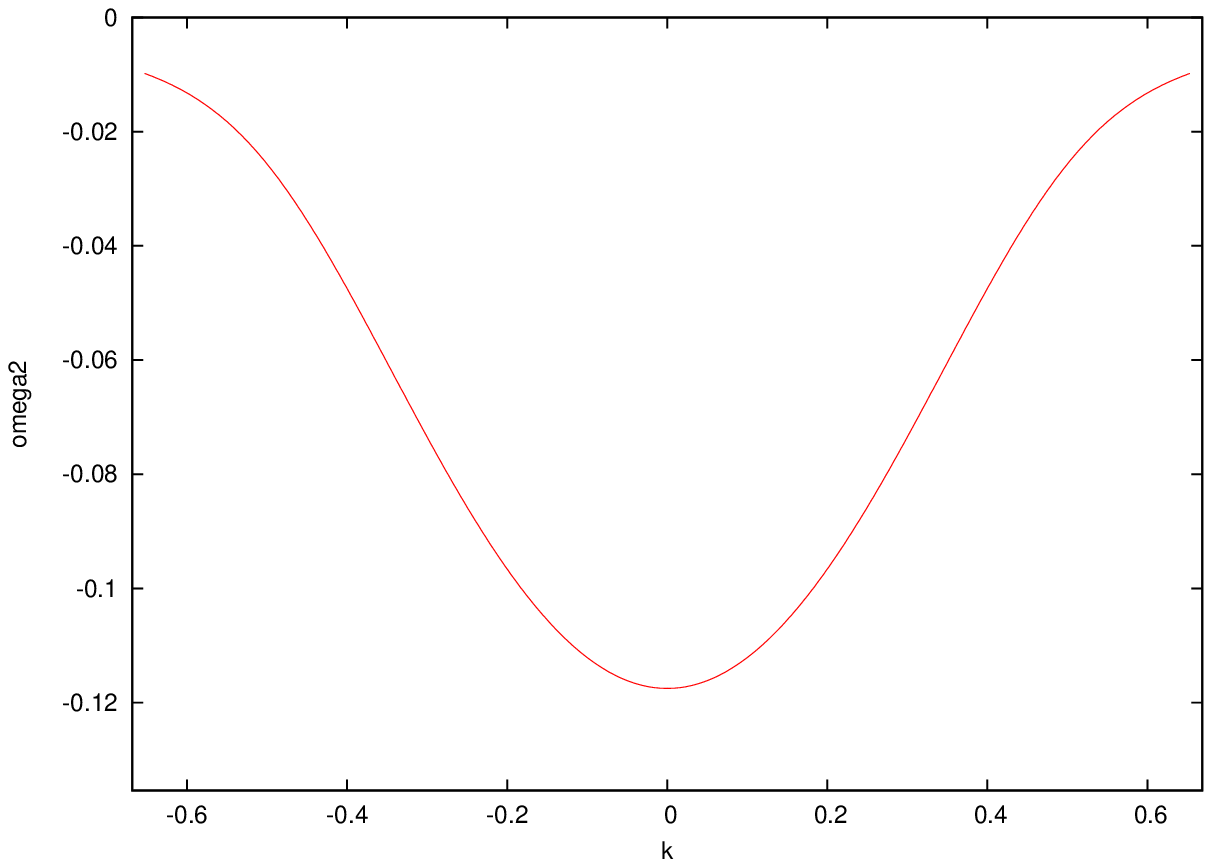}}
\hss}
\caption{The Jacobi determinant for Eqs.\eqref{S0} 
for various values of $\tilde{\kappa}$ (left) and  
the dispersion relation $\tilde{\omega}^2(\tilde{\kappa})$ (right) for the 
$n=\nu=1$, $\beta=2$, 
$\sin^2\thetaw=0.23$ vortex in the large current limit.  
}
\label{Fig_Large_I_Sector1}
\end{figure}

Introducing the five-component vector and the potential energy matrix  
\begin{equation}
\Psi=\left( \begin{array}{c} 
		X_1\\
		X_2\\
		X_3\\
		X_4\\
		X_5 
	\end{array}\right)	\, ,~~~~~~~~~~~
\mathcal{U}= \left( \begin{array}{ccccc} 
 M_1	&Q	&0	&0	&R	\\	
 Q	&M_2	&S	&S	&0	\\	
 0	&S	&M_+	&T	&U_+	\\	
 0	&S	&T	&M_-	&U_-	\\	
 R	&0	&U_+	&U_-	&M_0	\\	
\end{array}\right) 
\end{equation}
with the matrix elements 
\begin{align}
 M_1 &= \frac{V_3^2}{x^2}+U_1^2,~~~
 M_2 = \frac{1}{x^2}+U_1^2,~~~~
 M_\pm = \frac{(V_3\mp1)^2}{x^2}\pm\frac{2\partial_xV_3}{x}+U_1^2,~~~~
 M_0 = U_1^2	\, , \notag  \\
 Q &= -\sqrt{2}\partial_xU_1,~~~
 R =-\sqrt{2}S=-2{\tilde{\kappa}}U_1,~~~
 U_\pm = \sqrt{2}\left(\partial_xU_1\pm\frac{U_1V_3}{x} \right),~~~
 T = \frac{U_1^2}{2},	
\end{align}
the unstable sector is described by 
\be                  \label{S0}
-\frac{1}{x}\left(x\Psi^\prime_x\right)^\prime_x+\mathcal{U}\Psi=\Lambda\Psi,  
\ee
where $\Lambda=\tilde{\omega}^2-\tilde{\kappa}^2$.  
Fig.\ref{Fig_Large_I_Sector1} shows
the Jacobi determinant $\Delta(\rho)$ for various values of $\tilde{\kappa}$, and 
it seems that it always has a zero at some $\rho$, at least we could not find an
upper bound $\tilde{\kappa}_{\rm max}$ beyond which $\Delta(\rho)$ ceases to vanish. 
Since such a bound always exists for small currents, it 
should presumably exist also for large currents, 
but to find it one should probably refine the approximation \eqref{S0} 
to take into account the region outside the core. 
At present, it seems that the description \eqref{S0} is valid for any $\tilde{\kappa}$  
if ${\cal I}\to\infty$ or, if ${\cal I}$ is large but finite, up to 
some large but finite value of $\tilde{\kappa}$.

We then solve the eigenvalue problem \eqref{S0} looking for bound states 
with the boundary conditions 
 $X_3\sim X_5=O(1)$, 
 $X_1\sim X_2= O(x) $,
 $X_4=O(x^2)$
at small $x$, while at large $x$ 
\begin{align}\label{LARGE_I_ASYMPT_S1}
 X_1\pm X_5 &\sim X_3+X_4\mp\sqrt{2}X_2\sim
\exp\{-\int^x \sqrt{(U_1\mp \tilde{\kappa})^2-\tilde{\omega}^2}\,dx\}	\, , \notag \\
{X_3-X_4} &\sim \exp\{-\sqrt{\tilde{\kappa}^2-\tilde{\omega}^2}\,x\}	\,. 
\end{align}
This gives the dispersion relation $\tilde{\omega}^2(\tilde{\kappa})$ 
shown in Fig.\ref{Fig_Large_I_Sector1},
from where 
\be
\omega^2(\kappa)={\cal I}^2\tilde{\omega}^2({\kappa}/{\cal I}).
\ee
We see that the negative mode eigenvalue is large 
for large currents, $\omega\sim {\cal I}$, which relates  to the fact that the 
corresponding eigenmode is localized within a very short interval 
of size $\sim 1/{\cal I}$
inside the core.  
It is worth noting that the one-well shape of this dispersion relation 
is qualitatively similar to what is shown  in Fig.\ref{FigDISP}.
This suggests that the approximate description provided by Eqs.\eqref{S0}
is essentially correct.  
Since the ratio $\kappa/{\cal I}$
is small unless $\kappa$ is very large, one has 
\be                                    \label{largeom}
\omega^2(\kappa)\approx {\cal I}^2\tilde{\omega}^2(0)\approx -0.12\,{\cal I}^2,
\ee
where the numerical coefficient is calculated for  $n=\nu=1$, $\beta=2$, 
$\sin^2\thetaw=0.23$.

\section{Charged vortices \label{cvort}}

\begin{figure}[ht]
\hbox to \linewidth{ \hss
	\psfrag{omega2}{\large$\omega^2$}
	\psfrag{gp2}{\large$\sin^2\thetaw$}
	\psfrag{beta=2}{\large $\beta=2$}
	\psfrag{beta=3}{\large$\beta=3$}
	\psfrag{beta=5}{\large$\beta=5$}
	\resizebox{8cm}{2.5cm}{\includegraphics{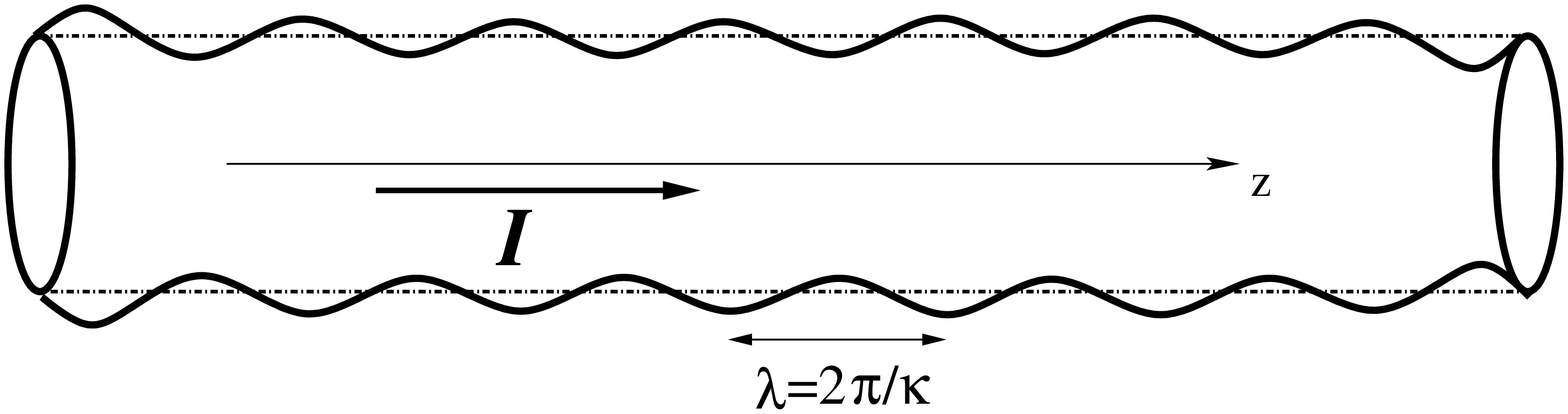}}
\hspace{4mm}
\resizebox{6cm}{5cm}{\includegraphics{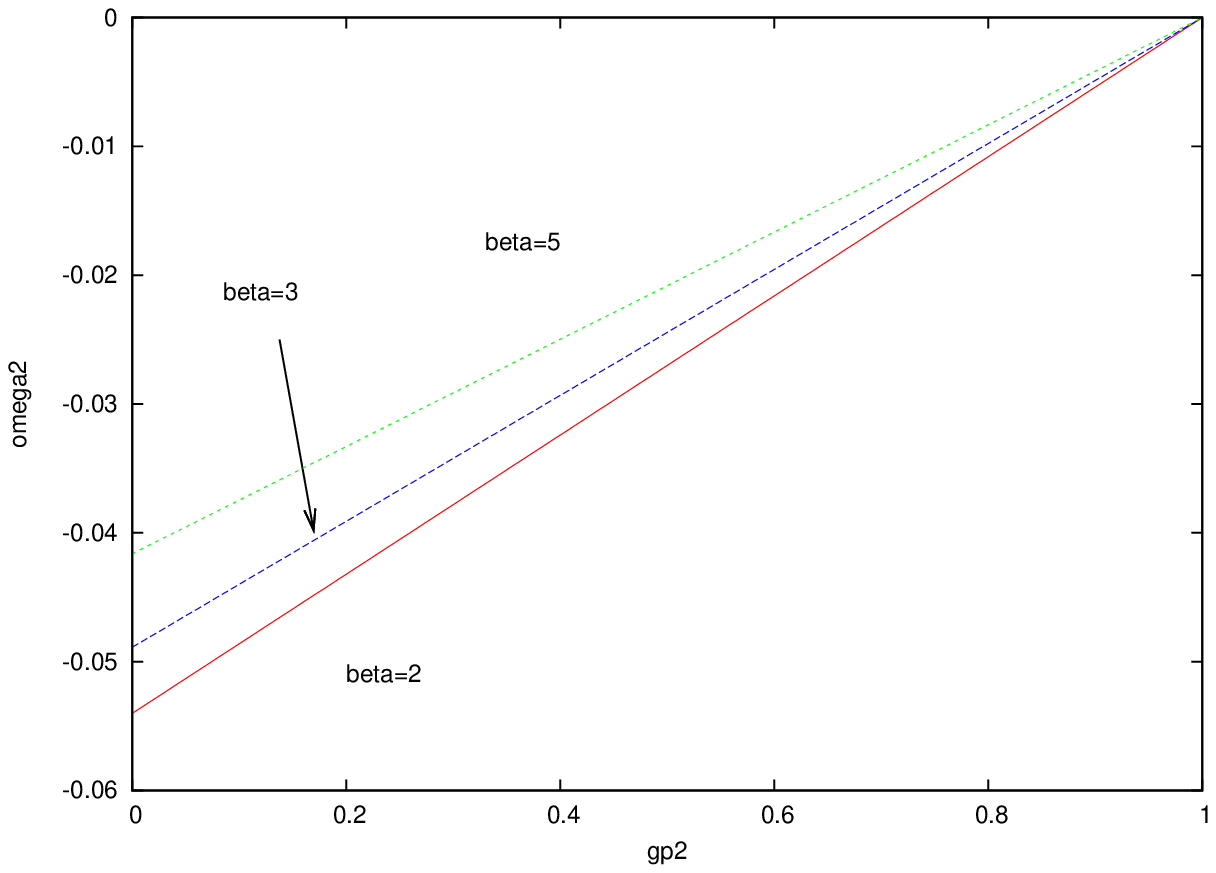}}

\hss}
\caption{\small The effect of a proper negative mode 
 on the vortex (left), and  the eigenvalue $\omega^2(0)$ of the homogeneous perturbation mode 
versus  $\thetaw$ for fixed $f_2(0)=0.1$ (right).
}
\label{cigar}
\end{figure}

Let us consider a bound state solution $\Psi_\kappa(\rho)$ of the 
Schr\"odinger problem \eqref{SCHRODINGER}  
 with the eigenvalue $\omega^2(\kappa)$ (setting for simplicity $m=0$). 
Injecting it into the mode decomposition 
\eqref{DECOMP} we  reconstruct the dependence  on all spacetime 
variables. The result will be a superposition of the real and imaginary parts of
\be						\label{eignm}
e^{i\,\phase(t,z)} \Psi_\kappa(\rho).
\ee
Here, 
 using \eqref{BOOST},   
\be							\label{bost0}
\phase(t,z)=\omega_b\,t+\kappa_b\,z
\ee
with $(\omega_b,\kappa_b)$ being the Lorentz-transformed (boosted) components of the spacetime 
vector $(\omega,\kappa)$,
\be						\label{bost}
{\omega}_b= \cosh(b)\,\omega+\sinh(b)\,\kappa,~~~~~~~~~~
{\kappa}_b=\cosh(b)\, \kappa+\sinh(b)\,\omega.
\ee
The boost parameter is related to the electric charge 
$I_0={\cal I}\sinh(b)$ (see \eqref{CURRENT}). 
Suppose that 
the mode under consideration is negative,  $\omega^2<0$, so that $\omega=i|\omega|$.
Then
\be					\label{QQQ}
\exp(i\,\phase)=\exp\left\{|\omega|(\cosh(b)t
+\sinh(b)z)\right\}
\exp\left\{i\kappa\,(\sinh(b)t
+\cosh(b)z)
\right\}.
\ee
Let us consider first the uncharged vortex, for which 
one has $b=I_0=0$ and so 
\be                         \label{QQQ1}
\exp(i\,\phase)=
\exp({|\omega|t})\exp({i\kappa z}),
\ee 
which grows in time but is periodic along $z$.
Let us call such negative modes {`proper'}.  
The effect of this instability   is schematically
shown in Fig.\ref{cigar} -- the vortex undergoes inhomogeneous, 
periodic in $z$  deformations 
which tend to segregate it into 
segments of length $\lambda=2\pi/\kappa$. 
Of course, this linear analysis is only valid as long as the perturbations are small, 
and so it does not imply that 
the vortex will actually break into segments -- 
such a possibility is unlikely in view of the current conservation. 
Since the current density becomes inhomogeneous,  
this  produces local inhomogeneities  in the electric charge distribution
in the form of a periodic sequence of positively and negatively charged regions
along the vortex.  

The generic vortex perturbation can be decomposed into a sum  
over eigenmodes. As time increases, one can expect  
this sum to become dominated by 
negative modes whose growth rate is maximal (provided that the perturbation remains small).
As is seen in Fig.\ref{FigDISP},   
$|\omega(\kappa)|$ is maximal for $\kappa=\pm\kappa_{\rm min}$
if ${\cal I}<{\cal I}_\star$ and for $\kappa=0$ if ${\cal I}>{\cal I}_\star$. 
Therefore, for small currents the vortex will probably tend to segregate into segments
of length $2\pi/\kappa_{\rm min}$ while for large currents it will rather expand 
homogeneously.

Since the wavevector of all negative modes is bounded, 
$|\kappa|<\kappa_{\rm max}$,  their wavelength is larger than
$2\pi/\kappa_{\rm max}$. 
Therefore, imposing periodic boundary conditions with period $2\pi/\kappa_{\rm max}$
will remove all these 
modes, because the vortex segment will not have enough 
room to accommodate them. 
Only the $\kappa=0$ will stay,
since it does not depend on $z$ and so can be considered as 
periodic with any period. This poses no problems if ${\cal I}=0$, 
or if ${\cal I}\neq 0$  but
$\thetaw=\pi/2$, since this mode is non-negative
in these cases. 
However,  
for $\thetaw<\pi/2$  and ${\cal I}\neq 0$
the homogeneous  mode is negative. This 
is seen in Fig.\ref{FigDISP}, in Table I, and also in Fig.\ref{cigar} which shows 
that $\omega^2(0)$ is negative for ${\cal I}\neq 0$ and vanishes only for $\thetaw=\pi/2$.
This means that the generic vortices cannot be stabilized by periodic boundary conditions.

We did not find any simple arguments explaining why $\omega^2(0)$ should generically be negative. 
 Since $\omega^2(0)=0$ 
for $\thetaw=\pi/2$, when the massless fields decouples, one can suspect that the 
explanation could be related to the presence of the long-range field in the system.  
However, the massless fields decouple also for $\thetaw=0$, but in this case one has  $\omega^2(0)<0$ 
(see Fig.\ref{cigar}). It is therefore likely that the explanation should rather be related to the
non-Abelian nature of the background solutions. Indeed, the non-linear commutator terms 
are generically present in the backgrounds, but they vanish  for 
$\thetaw=\pi/2$ (when the SU(2) field decouples) or for ${\cal I}=0$ (because Z strings are 
embedded Abelian solutions), that is exactly when  $\omega^2(0)$ vanishes.

We have  tried to analytically evaluate $\omega^2(0)$ for small currents by using the method 
applied in 
Ref.~\cite{FORGACS}. In this method both the background and perturbation equations are expanded
in powers of the small parameter $q=f_2(0)$ and then solved order by order. 
We have found that $\omega^2(0)=-cq^2+\ldots$ where $c>0$, so that the homogeneous mode becomes
negative for however small currents. It is also negative for large currents, 
as is shown by \eqref{largeom}. 
Therefore, it is generically negative.

\begin{figure}[ht]
\hbox to \linewidth{ \hss
	\psfrag{y}{}
	\psfrag{lnx}{$\ln(1+x)$}
	\resizebox{7cm}{5cm}{\includegraphics{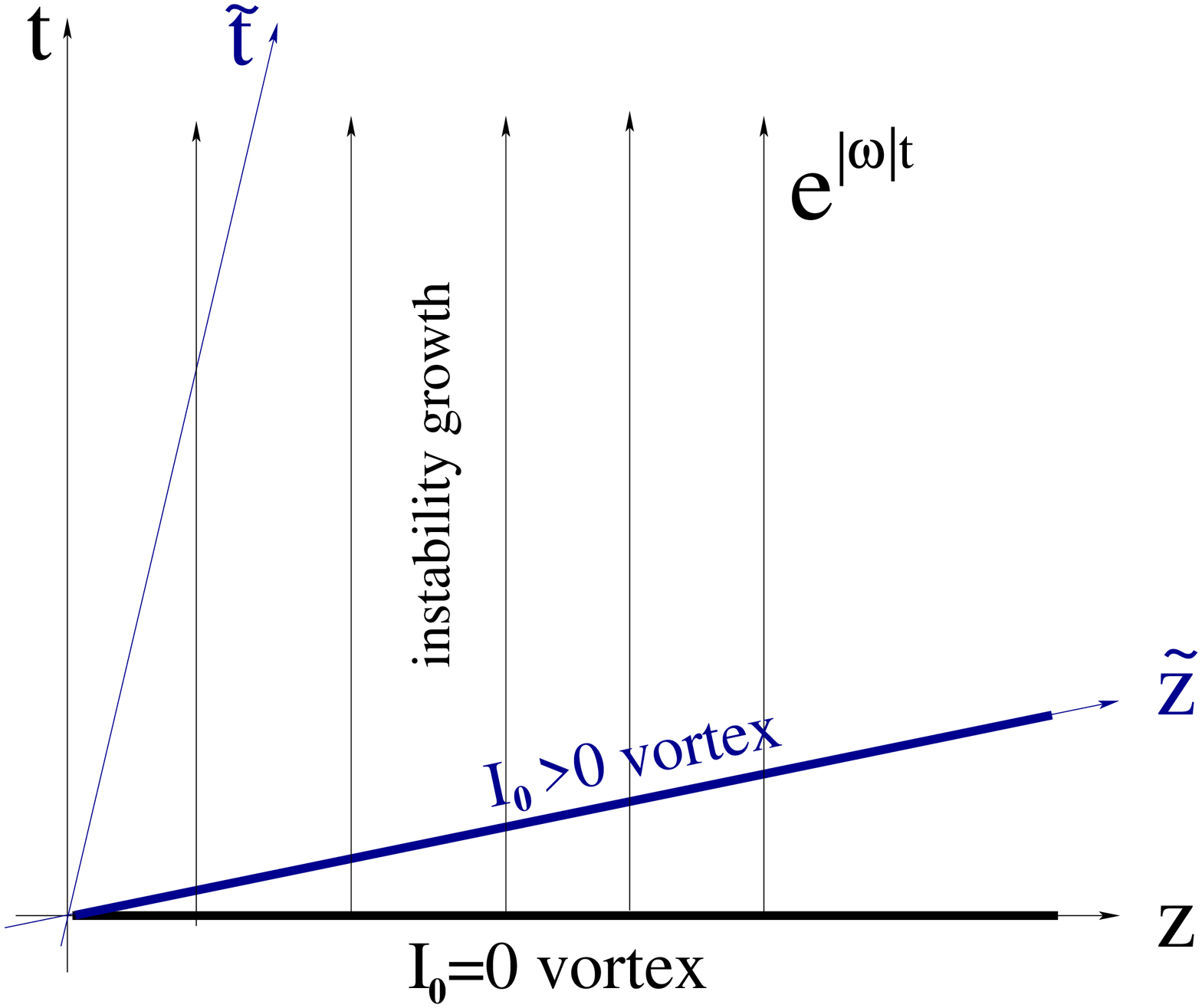}}
\hspace{4mm}
\resizebox{8cm}{4cm}{\includegraphics{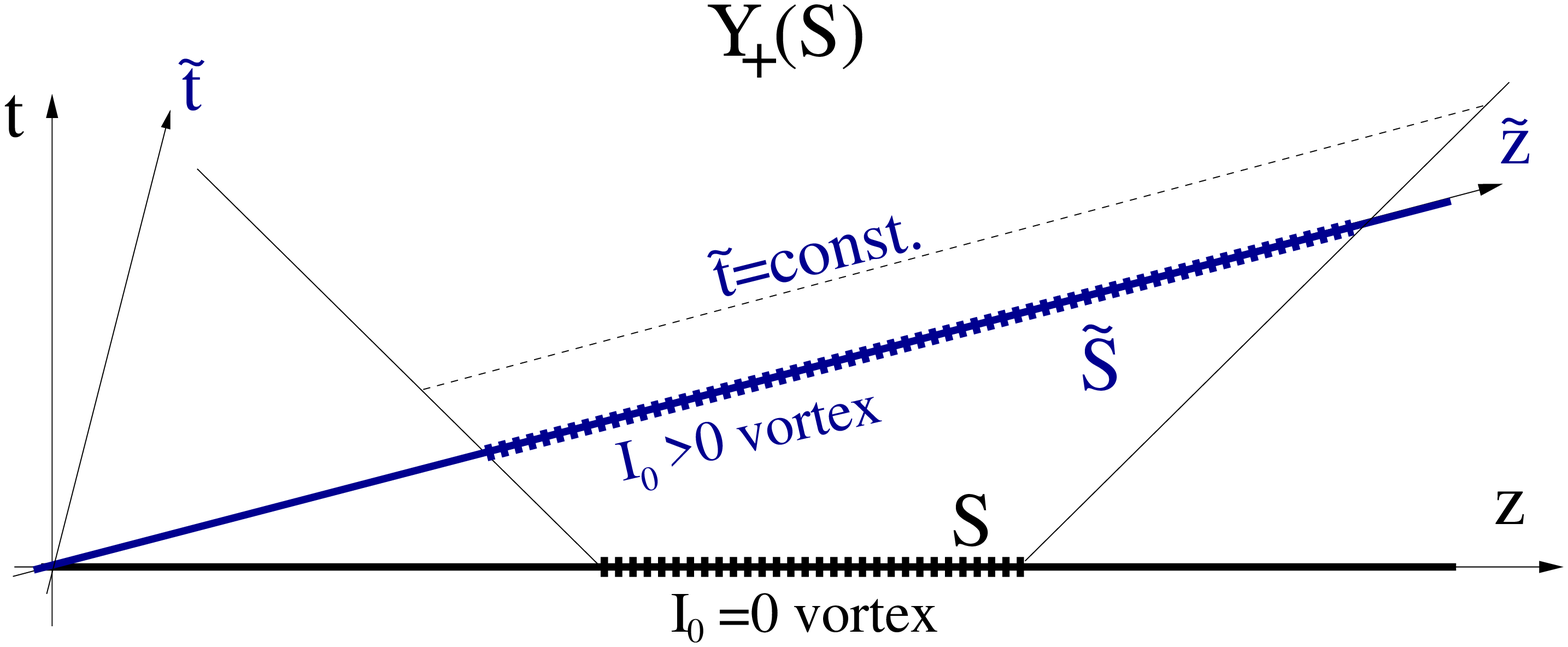}}

\hss}
\caption{\small Left: The proper negative modes grow along the spacetime history lines.
The restframe time of the $I_0=0$ vortex flows in the same direction, while for the 
boosted vortex the time direction is different so that the instability grows
not only with time $\tilde{t}$ but also with $\tilde{z}$. Right: Spacetime evolution
of the initial data with compact support $S$.  
}
\label{boost}
\end{figure}

Let us now consider electrically charged vortices with $I_0\neq 0$. 
Since they can be obtained by boosting the $I_0=0$ vortex,
their perturbation can also be obtained in the same way. 
Boosting the proper modes \eqref{QQQ1} gives 
the negative modes \eqref{QQQ} of the charged vortex, and 
we shall call such modes `boosted' in order to distinguish them from the proper modes. 
The boosted modes grow not only in time 
but also in space, along the vortex, which  is a simple 
consequence of the fact that the time/space directions for the 
boosted vortex are not the same as for the $I_0=0$ vortex. 
The proper negative modes of the latter grow only in time when considered in the 
restframe, but the observer comoving with the charged vortex will see   
the very same modes  
grow not only in time but also in space  (see Fig.\ref{boost}). Equivalently, 
one can say that  
the boost renders complex the wavevector $\kappa_b$ in \eqref{bost},
since $\omega$ is imaginary.
% and so the boosted negative modes 
%grow with $z$ as $\exp(\pm i{\kappa}_b z)$. 

Since the boosted negative modes grow with $z$, they can  be used only within a finite
range of $z$ for the small perturbation theory to be valid. This can be achieved by forming 
 wavepackets. 
Let us consider a wavepacket of the proper eigenmodes of the 
 $I_0=0$ vortex, 
\be                                     \label{evol}
\delta f(t,\rho,z)=\int d{\kappa}\, C(\kappa) 
e^{i\omega(\kappa)t+i\kappa z}\Psi_\kappa(\rho)+\ldots
\ee
where
the dots stand for the contribution of the scattering states (solutions 
of \eqref{SCHRODINGER} which do not vanish at infinity).
Assuming  the initial perturbation
 $\delta f(0,\rho,z)$ to have a compact 
support $S$ along $z$-axis, its time evolution  
will be contained within the  
spacetime domain $Y_{+}(S)$ 
causally connected with
$S$  (see Fig.\ref{boost}). 
By simply changing the coordinates,
$t=\cosh(b)\tilde{t}+\sinh(b)\tilde{z}$  
and 
$z=\cosh(b)\tilde{z}+\sinh(b)\tilde{z}$,
the same wave packet can be reexpressed as sum over boosted 
modes,
\be                                     \label{evol11}
\delta{f}(\tilde{t},\rho,\tilde{z})=
\int d{\kappa}\, C(\kappa) 
e^{i\,\phase(\tilde{t},\tilde{z})} \Psi_\kappa(\rho)
+\ldots 
\ee
where $\phase$ is given by \eqref{bost0}--\eqref{QQQ}. 
If there are negative modes in \eqref{evol}, 
then \eqref{evol11} will contain growing in 
$\tilde{z}$ terms, but since $\tilde{z}$ actually varies only within the 
finite range inside $Y_{+}(S)$  for a fixed $\tilde{t}$, the whole sum is 
bounded. One can therefore view this wavepacket  as perturbation of the charged vortex 
with the  initial distribution $\delta {f}(0,\rho,\tilde{z})$ contained in $\tilde{S}$ 
(see Fig.\ref{boost}). 
If $\delta f(t,\rho,z)$ grows with $t$ then 
$\delta{f}(\tilde{t},\rho,\tilde{z})$ will grow with $\tilde{t}$,
hence if the $I_0=0$ vortex is unstable then so is the $I_0\neq 0$ vortex.

So far, however, the symmetry between the $I_0=0$ and $I_0\neq 0$ vortices is incomplete, 
because we have found only the proper negative modes for the former 
and only the boosted negative modes for the latter. 
These modes were  obtained by solving the radial equations \eqref{SCHRODINGER} with 
real ${\kappa}$ and real $\omega^2<0$ and they are spatially periodic 
in the vortex restframe but  become non-periodic 
after the boost. One might therefore  think that periodic boundary conditions 
could stabilize the charged vortices, since they will remove all boosted modes.  
However, there could be also solutions of Eqs.\eqref{SCHRODINGER} 
giving rise to negative modes which are initially non-periodic 
but become periodic after the boost.
The boosted value $\kappa_b$
in \eqref{bost} should then be real, hence 
one should look for bound state solutions of 
\eqref{SCHRODINGER} for complex parameters
\be                                            \label{compl}
\omega=\gamma-i\Omega,~~~~~\kappa=K+i\,\Omega\tanh(b), 
\ee  
where $\gamma=\gamma(b,K)$, 
$\Omega=\Omega(b,K)$. It is worth noting that a similar 
recipe was considered 
within the stability analysis of the boosted black strings 
in the theory of gravity \cite{Branes}.  
Inserting this in \eqref{bost},
the imaginary part of $\kappa_b$ vanishes
and one obtains 
\be					\label{QQQ2}
\exp(i\,\phase)=\exp\left(\Omega_b t\right)
\exp\left(i\gamma_b t
+i \kappa_b z
\right)
\ee
with 
\be
\Omega_b=\frac{\Omega t}{\cosh(b)},~~~~~
\gamma_b=\cosh(b)\gamma+\sinh(b)K,~~~~
\kappa_b=\cosh(b) K+\sinh(b)\gamma. 
\ee
Since $\exp(i\,\phase)$ grows in time and 
has the harmonic $z$-dependence, 
this corresponds to proper negative modes of the charged
vortex with $I_0={\cal I}\sinh(b)$. 

\begin{figure}[ht]
\hbox to \linewidth{ \hss
	\psfrag{P}{\large$\kappa_b$}
	\psfrag{Omega_b}{\large$\Omega_b$}
	\psfrag{gamma_b}{\large$\gamma_b$}
	\resizebox{8cm}{6cm}{\includegraphics%{Dispersion_cplx2_s=0.7_vs_P_boost=0_to_0.4.eps}}
{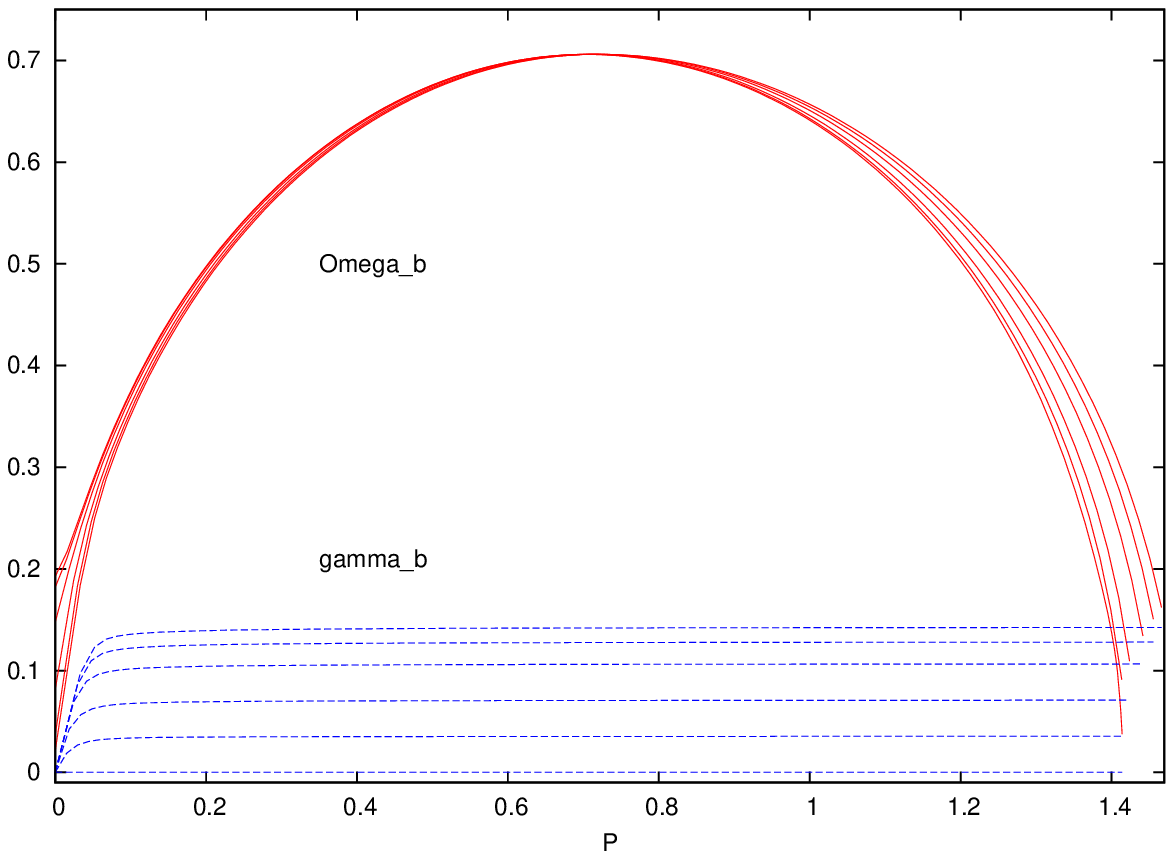}}
\hspace{4mm}
	\psfrag{Omega_s=0.7}{\large$\Omega,~~{\cal I}=0.08$}
	\psfrag{Omega_s=0.35_over_4}{\large$0.25\times \Omega,~~{\cal I}=2.67$}
	\psfrag{boost}{\large$b$}
\resizebox{8cm}{6cm}{\includegraphics{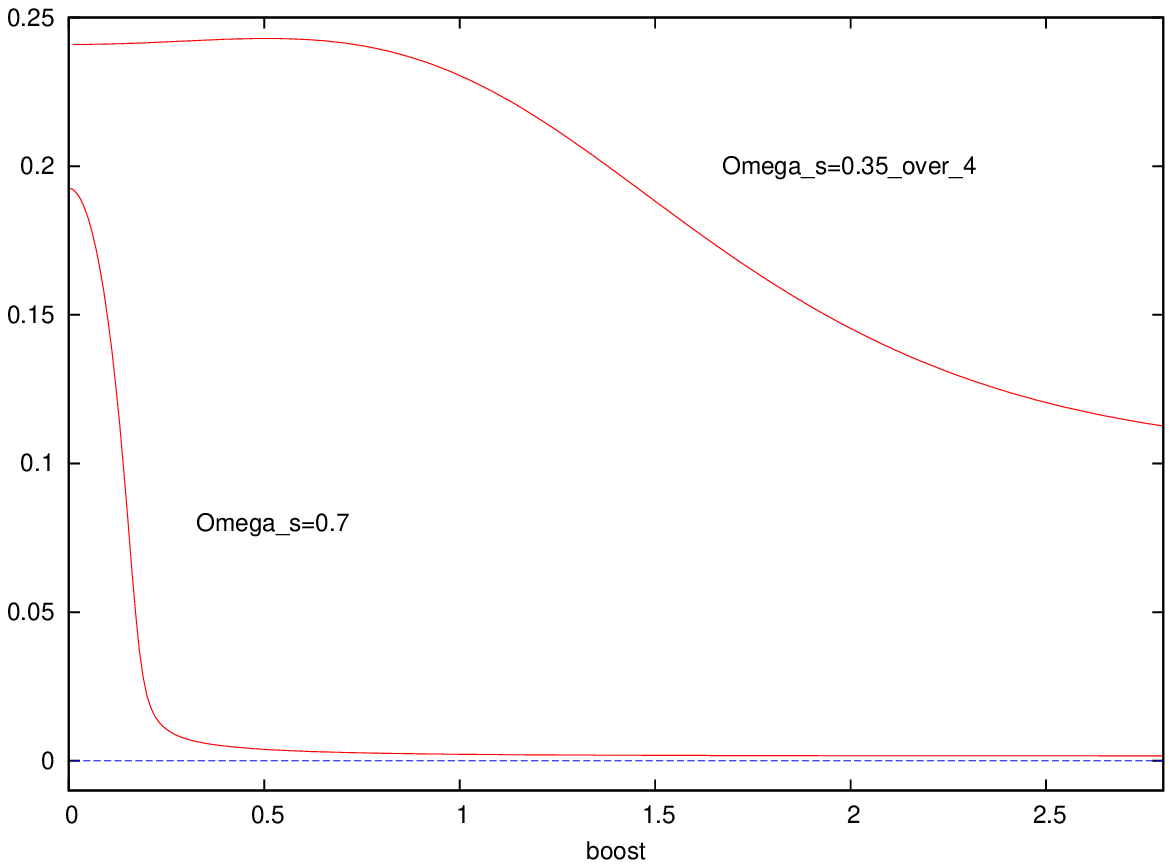}}

\hss}
\caption{\small  Left: Real and imaginary frequency parts 
$\gamma_b$ and $\Omega_b$ versus $\kappa_b$ for 
perturbations of the boosted vortex with ${\cal I}=0.08$  
 for 
several values of the boost $b\in[0,0.4]$.  
Right:  $\Omega$ versus the boost parameter $b$ for the homogeneous ($\kappa_b=0$) 
perturbation mode for the boosted vortices with ${\cal I}=0.08$  and ${\cal I}=2.67$. 
In both panels   $\beta=2$, $\sin^2\thetaw=0.23$, $n=\nu=1$, $m=0$. 
}
\label{complx}
\end{figure}
 
The next question is whether solutions of
Eqs.\eqref{SCHRODINGER} for the complex parameter values 
\eqref{compl} exist.  If 
$b=0$ then $\kappa$ is real and 
we should recover the already known solutions 
with real 
$\omega^2=-\Omega^2$, therefore one has $\gamma=0$ in this case. 
If $b\neq 0$ then both $\omega^2$ and $\kappa$ become complex
so that equations  \eqref{SCHRODINGER} should be complexified 
as well. We therefore obtain a system of $16$ linear complex equations,
which is equivalent to $32$ real equations. 
This does not  mean  that the number of degrees of freedom doubles,
since even for $b=0$ we actually had $32$(=40-8) real equations split 
into two independent subsystems of $16$ equations each, equivalent
to each other upon \eqref{REPLACE}. For $b\neq 0$
we still have the $32$ equations, but they no longer split
into two  subsystems. 

Solving numerically the $32$ coupled equations is considerably 
more time consuming than solving the $16$ equations. This is why 
we did not 
carry out a systematic analysis of the parameter space but studied
instead just several representative cases.  
We integrated the equations looking for bound state solutions 
with the boundary conditions given by 
the complexified version 
of \eqref{BC_AXIS},\eqref{BC_ASYMPT}. Choosing a value of $b$, 
we have managed to explicitly construct such solutions and determine 
$\gamma$ and $\Omega$ as functions of $K$. 

It turns out that the dispersion relation for $\Omega_b$ 
 against $\kappa_b$  remains
 qualitatively the same for $b\neq 0$
as for $b=0$ (see Fig.\ref{complx}).  
If the current ${\cal I}$ is small and $b$ is fixed then $\Omega(\kappa_b)$
starts at a non-zero value at $\kappa_b=0$, increases and reaches maximum,
then decreases and vanishes for some maximal value $\kappa_b=\kappa_{\rm max}$
(see Fig.\ref{complx}). For large currents $\Omega(\kappa_b)$
 decreases monotonously from its
value  at $\kappa_b=0$ till zero.  
One has $\Omega(\kappa_b)=\Omega(-\kappa_b)$.

\begin{figure}[ht]
\hbox to \linewidth{ \hss
	\psfrag{y}{}
	\psfrag{lnx}{$\ln(1+x)$}
	\resizebox{8cm}{5cm}{\includegraphics{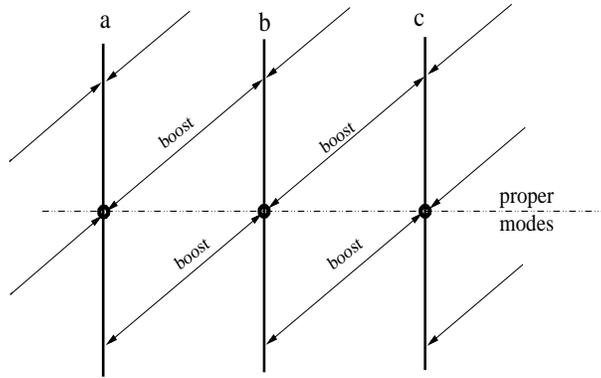}}

\hss}
\caption{\small The set of negative modes for any given vortex charge, for example  
$I_0={\cal I}\cosh(b)$,   
is represented by the vertical line. The proper modes are represented
by the fat points. 
There is one-to-one correspondence  between 
modes for different $I_0$ via boosts. 
}
\label{proper}
\end{figure}

The proper negative modes therefore exist for any value of charge and not only for $I_0=0$. 
To completely restore the symmetry between solutions with different $I_0$,
we note that the proper negative modes for any  $b$
can be boosted towards  a different value 
 of the boost parameter, $B$, say. 
This will give boosted negative modes of the $I_0={\cal I}\sinh(B)$ vortex 
proportional to
\be                               \label{B}
\exp\left\{\Omega_b\,{\cosh(B-b)}\,t+\Omega_b\,{\sinh(B-b)}\,z\right\}
\exp\left\{i(\gamma_B t+\kappa_B z) \right\}
\ee 
with $\kappa_B=\cosh(B)K+\sinh(B)\gamma$ and 
$\gamma_B=\cosh(B)\gamma+\sinh(B)K$. 
Therefore, for any vortex charge $I_0={\cal I}\sinh(B)$ 
there are proper negative modes, but also infinitely many  
boosted  modes labeled by $b\neq B$.
The space of negative modes has  the same structure for any value of charge, since 
 there is  one-to-one  correspondence  
 between modes for different charges via boosts, as schematically shown in 
Fig.\ref{proper}.

The boosted negative modes are  non-periodic in space and can contribute 
only to the instability of infinitely long vortices, but they will be removed
by imposing on vortex the periodic boundary conditions. The proper modes will stay then,
but if the period is less than $2\pi/\kappa_{\rm max}$ then 
they will be removed as well, apart from the $\kappa_b=0$  mode. 
We know that for $I_0=0$ this mode is generically negative, but perhaps things may change
for $I_0\neq 0$~? 
We therefore trace $\Omega$ for this mode against $b$
and find that it decreases very rapidly with $b$, especially for small currents 
(see Fig.\ref{complx}), so that the instability growth rate decreases when the vortex
charge $I_0$ increases. However,
it is not clear from these data if  $\Omega$ always stays finite or
eventually vanishes at some large value of $b$.  
It seems however that the latter option is impossible, 
since $\kappa_b=0$ implies that $\gamma=K=0$. 
Setting $\Omega=0$ would therefore mean that $\omega=\kappa=0$,
but since $\kappa=0$ is real, this solution should be contained in the  
previously obtained dispersion relation $\omega^2(\kappa)$. 
However, we know from the previous analysis that $\omega^2\neq 0$
for $\kappa=0$ (unless ${\cal I}=0$), and so the value $\Omega=0$ is impossible. 
Therefore, there is no critical value of boost for which the homogeneous instability 
would disappear.

\section{Conclusions}

We study  in this paper the stability of the 
superconducting vortex solutions in the Weinberg-Salam theory 
described in Ref.~\cite{JGMV2}. 
Such vortices are characterized by
 a constant  electric current $I_3={\cal I}\cosh(b)$ and 
linear electric charge density $I_0={\cal I}\sinh(b)$ 
comprising  a spacelike  vector $(I_0,I_3)$. 
Fixing  ${\cal I}$, vortices with different
values of the charge $I_0$ can be related 
to each other by Lorentz boosts,
in particular there exists the restframe where $I_0=0$.  
For ${\cal I}\to 0$ all solutions become Z strings, 
while for $\thetaw\to\pi/2$ and $\beta>1$ 
they reduce to the twisted semilocal strings studied in Ref.~\cite{SL}.

We consider generic vortex perturbations in the linear approximation
and find that after separating the variables the 
perturbation equations reduce to the effective $16$-channel 
Schr\"odinger problem
 \eqref{SCHRODINGER}. 
This problem admits bound state solutions 
with $\omega^2<0$ whose dispersion relation $\omega^2(\kappa)$ is 
shown in Fig.\ref{FigDISP} and
tabulated in Table I. These solutions describe the `proper' negative modes
of the $I_0=0$ vortex.   
Choosing the parameters $\omega,\kappa$
in Eqs.\eqref{SCHRODINGER}
to be complex
gives bound state solutions describing 
proper negative modes
of the charged vortices.  
As a result, for any given value of charge $I_0$ there is a one-parameter
family of proper negative modes which can be labeled by the  
wavevector $\kappa_b$. 
These perturbation modes grow in time favoring  segregation 
of the homogeneous vortex into segments,
although one cannot conclude from the perturbative analysis whether it will actually  
break in pieces in the long run. 

Since vortices with different $I_0$ are related by Lorentz boosts, their perturbations
can be related in this way too. Boosting the proper negative modes of the  
$I_0={\cal I}\sinh(b)$ vortex one obtains negative 
modes of the $I_0={\cal I}\sinh(B)$ vortex, so that the latter acquires in fact 
an additional
infinity of negative modes labeled by $b\neq B$. These
`boosted'  modes grow with $z$ but they 
 can form localized wavepackets to contribute to the instability of 
infinitely long vortices. Since they are non-periodic in space,  
they can be removed by imposing
periodic boundary conditions along the vortex. 

The proper negative modes are proportional to 
$
%\exp\left(\Omega_b(\kappa_b) t\right)
%\exp\left(i\gamma_b(\kappa_b) t
\exp\{{i \kappa_b z}\}
$
and so they can be made compatible with the periodicity along $z$ 
by adjusting the value of $\kappa_b$. 
However, they exist only for 
$|\kappa_b|<\kappa_{\rm max}$ and so choosing the period to be less than 
$2\pi/\kappa_{\rm max}$ 
the vortex segment will not have enough room to accommodate these modes. 
All of them will therefore be removed, apart from the 
$\kappa_b=0$ 
mode which 
is independent of $z$ and can be considered as periodic with any period. 
Therefore, the only remaining vortex instability is associated with this homogeneous mode. 

In some cases one has $\omega=0$ for $\kappa_b=0$, as for example for 
$\thetaw=\pi/2$ and for any ${\cal I}$ (semilocal vortices),
 or for ${\cal I}=0$ and for any $\thetaw$
(Z strings). In these cases the  homogeneous mode is not negative and so 
 the short periodic vortex segments turn out to be stable. 
In particular, Z strings  
 can be stabilized in this way by 
 passing to the gauge \eqref{003aZ1} and then imposing the
periodic boundary conditions which break  the gauge invariance. 
However, in the generic case the homogeneous mode is negative
and it renders the vortex unstable with respect to the homogeneous expansion 
even after imposing periodic boundary conditions. 

At the same time, it is possible  that the homogeneous negative mode could  be removed by 
the curvature effects.  Specifically, let us suppose that one `cuts out' 
a finite vortex segment, bends it and identifies 
its extremities  to make a loop. Then, since 
the loop thickness cannot be larger than its radius, 
the homogeneous expansion of the vortex segment should inevitably stop at some point.
Therefore, the homogeneous instability will be removed, suggesting that     
the loop could be stable. Of course, this argument
is only qualitative. Moreover, new instabilities 
could appear  when bending  the vortex. 
However, any possibility to have stable electroweak solitons, as for example vortex
loops, could be very important.
 Such loops could be balanced against contraction by the centrifugal 
force arising from the momentum circulating along them.
Since momentum flows along vortices with $I_0\neq 0$, they can be 
naturally used to `make' the loops.  
All this suggests that spinning vortex loops -- electroweak analogs 
  of the `cosmic vortons' \cite{Davis} -- could 
exist and could perhaps even be stable. 
Of course, verification of this conjecture 
requires serious efforts, since so far vortons
have been explicitly constructed  only in a simple
scalar field model \cite{RV2008}, \cite{BS2009}. However, 
if the electroweak vortons indeed 
exist and are stable, they could be a dark matter candidate.

There could be other physical manifestations of the superconducting vortices.
They 
could perhaps be created either at high temperatures
or in high energy collisions, and since they are non-topological, they could exist in the 
form of finite segments. If their extremities are attached to something (charged clouds),
then they could be spatially periodic and 
transfer
charge between different regions of space like  `electroweak thunderbolts'.  
Non-periodic vortex segments 
should decay emitting jets of $W^{\pm}$ through its extremities,
which could perhaps be detectable at the LHC. Specifically, large magnetic fields
similar to those inside the vortex and also large currents can be created in the LHC  
heavy ion collisions.  This could lead to creation of virtual vortex segments whose 
subsequent  disintegration would be accompanied by showers of $W^{\pm}$'s. 
As a result, if one observes an excessive $W^{\pm}$ production in the collisions,
this could indicate the vortex segment creation. A similar way to detect the presence 
of the non-perturbative electroweak 
structures in the LHC collisions was discussed in \cite{AO}.

\section*{ACKNOWLEDGEMENTS}
We would like to thank Jan Ambjorn, Christos Charmousis, 
Maxim Chernodub, Tom Kibble, Frans Klinkhamer, Alexey Morozov, Niels Obers, 
Paul Olesen, Eugen Radu, Mikhail Shaposhnikov, 
Sergey Solodukhin, Toby Wiseman, 
and Andreas Wipf for discussions and remarks at various 
stages of this work.

\renewcommand{\thesection}{APPENDIX A}
\section{Background field equations}
\renewcommand{\theequation}{A.\arabic{equation}}
\setcounter{equation}{0}
\setcounter{subsection}{0}

With the parametrization \eqref{003} the field equations 
\eqref{P0}--\eqref{P2} 
reduce to  two U(1) equations 
(with $^\prime\equiv \frac{d}{d\rho}$)
\begin{align}                                           \label{ee1}
\frac{1}{\rho}(\rho\Y')'&=\left.\left.\frac{g^{\prime\,2}}{2}
\right\{(\Y+\Om_3)\f^2+2\,\Om_1^{}\f^{}\p^{}+(\Y-\Om_3)\p^2\right\},
\\
\rho\left(\frac{\Z^\prime}{\rho}\right)^\prime&=\left.\left.\frac{g^{\prime\,2}}{2}
\right\{(\Z+\W_3)\f^2+2\,\W_1^{}\f^{}\p^{}+(\Z-\W_3)\p^2\right\},
\label{ee2}  
\end{align}
two Higgs equations 
\begin{align}                                             \label{ee3}
\frac{1}{\rho}(\rho\f^\prime)^\prime&=
\left\{\frac{\om^2}{4}\left[(\Y+\Om_3)^2+\Om_1^2\right]
+\frac{1}{4\rho^2}\left[(\Z+\W_3)^2+\W_1^2\right]
+\frac{\beta}{4}(\f^2+\p^2-1)
\right\}\f\nonumber \\
&+\left(\frac{\om^2}{2}\,\Y\Om_1+\frac{1}{2\rho^2}\,\Z\W_1\right)\p,
\\
\frac{1}{\rho}(\rho\p^\prime)^\prime&=
\left\{\frac{\om^2}{4}\left[(\Y-\Om_3)^2+\Om_1^2\right]
+\frac{1}{4\rho^2}\left[(\Z-\W_3)^2+\W_1^2\right]
+\frac{\beta}{4}(\f^2+\p^2-1)
\right\}\p  \nonumber \\
&+\left(\frac{\om^2}{2}\,\Y\Om_1+\frac{1}{2\rho^2}\,\Z\W_1\right)\f,\label{ee4}                                                   
\end{align}
four Yang-Mills equations 
\begin{align}
\frac{1}{\rho}(\rho\Om_1^\prime)^\prime
&=
-\frac{1}{\rho^2}\left(\W_1\Om_3-\W_3\Om_1\right)\W_3+\frac{g^2}{2}
\left[\Om_1(\f^2+\p^2)+2\Y\f\p\right],            \label{ee5}   \\
\frac{1}{\rho}(\rho\Om_3^\prime)^\prime
&=
+\frac{1}{\rho^2}\left(\W_1\Om_3-\W_3\Om_1\right)\W_1 + \frac{g^2}{2}
\left[(\Om_3+\Y)\f^2+(\Om_3-\Y)\p^2\right],       \label{ee6}\\
\rho\left(\frac{\W_1^\prime}{\rho}\right)^\prime
&=
+\om^2 \left(\W_1\Om_3-\W_3\Om_1\right)\Om_3+\frac{g^2}{2}
\left[\W_1(\f^2+\p^2)+2\Z\f\p\right],             \label{ee7}  \\
\rho\left(\frac{\W_3^\prime}{\rho}\right)^\prime
&=
-\om^2 \left(\W_1\Om_3-\W_3\Om_1\right)\Om_1+\frac{g^2}{2}
\left[(\W_3+\Z)\f^2+(\W_3-\Z)\p^2\right],         \label{ee8}
\end{align}
and a first order constraint
\be                                     \label{CONS1}
\sigma^2( \Om_1^{}\Om_3^\prime-\Om_3^{}\Om_1^\prime)
+\frac{1}{\rho^2}\,(\W_1^{}\W_3^\prime-\W_3^{}\W_1^{\prime})-
g^2(\f^{}\p^{\prime}-\p^{}\f^{\prime})=0.
\ee

\renewcommand{\thesection}{APPENDIX B}
\section{Perturbation equations \label{APP_PERT}}
\renewcommand{\theequation}{B.\arabic{equation}}
\setcounter{equation}{0}
\setcounter{subsection}{0}

Fixing the gauge, decoupling 
the ghost modes as described in the main text
and using the parametrization \eqref{lincomb} 
for perturbations in the physical sector,  the perturbation equations 
can be written in the form of    
the $16$-channel  Schr\"odinger problem 
\be                  \label{SO}
-\frac{1}{\rho}\left(\rho\Psi^\prime\right)^\prime+\mathcal{U}\Psi=\omega^2\Psi,  
\ee
where the $16$-component vector $\Psi$ and the $16\time 16$ 
symmetric potential matrix $\mathcal{U}$ read
\begin{align}\label{PSI_U}
\Psi&=\left( \begin{array}{c} 
		\vec{\mathcal{Z}}\\
		\vec{\mathcal{A}}\\
		\vec{\mathcal{W}}^+\\
		\vec{\mathcal{W}}^-\\
		\vec{\mathcal{H}} 
	\end{array}\right)	\, ,&
\mathcal{U}&=\left( \begin{array}{ccccc} 
\DZ	&\Gza	&\Gzwp	&\Gzwm	&\Gzh	\\	
\Gza	&\DA	&\Gawp	&\Gawm	&\Gah	\\	
\Gzwp	&\Gawp	&\DWp	&\Gww	&\Gwph	\\	
\Gzwm	&\Gawm	&\Gww	&\DWm	&\Gwmh	\\	
\Gzh	&\Gah	&\Gwph	&\Gwmh	&\DHH		
\end{array}\right)	\, .
\end{align}
Here 
\begin{align}\label{VECTOR_DEF}
\vec{\mathcal{Z}}&=\left(\begin{array}{c} 
\Zz\\
\Zp\\
\Zm
\end{array}\right)	\, , &
\vec{\mathcal{A}}&=
\left(\begin{array}{c} 
\Az\\
\Ap\\
\Am
\end{array}\right)	\, , &
\vec{\mathcal{W}^\pm}&=
\left(\begin{array}{c} 
\Wpmz\\
\Wpmp\\
\Wpmm
\end{array}\right)	\, , &
\vec{\mathcal{H}}&=
\left(\begin{array}{c} 
\Hop\\
\Hom\\
\Htp\\
\Htm
\end{array}\right)	\, ,
\end{align}
and $\DZ=\text{diag}\left(\Dzz,\Dzp,\Dzm\right)$ also 
$\DA=\text{diag}\left(\Daz,\Dap,\Dam\right)$ while 
\begin{align}\label{MATRIX_DIAG_DEF}
\DWpm&=\left( \begin{array}{ccc} 
\Dwpmz	&\pm\mathcal{Q}	&\pm\mathcal{Q}	\\	
\pm\mathcal{Q}	&\Dwpmp	&0	\\	
\pm\mathcal{Q}	&0	&\Dwpmm	\\	
\end{array}\right)	\, , &
\DHH&=\left( \begin{array}{cccc} 
\Dhop	&V_1	&V_+	&V_0	\\	
V_1	&\Dhom	&V_0	&V_-	\\	
V_+	&V_0	&\Dhtp	&V_2	\\	
V_0	&V_-	&V_2	&\Dhtm		
\end{array}\right) %	\, , \notag \\
\end{align}
with ($\eta=0,\pm 1$) 
\begin{align}\label{OPERATOR_DIAG}
\Dze ~&	= \frac{g^2v_1^2+\left(m-\eta\right)^2}{\rho^2}
	+g^2\sigma^2u_1^2+\kappa^2+\frac{1}{2}(f_1^2+f_2^2)-2g^2g^{\prime2}f_2^2	\, ,\notag \\
\Dae ~&	= \frac{g^{\prime2}v_1^2+\left(m-\eta\right)^2}{\rho^2}\
	+g^{\prime2}\sigma^2u_1^2+\kappa^2+2g^2g^{\prime2}f_2^2	,\notag \\
\Dwpme &	= \frac{v_1^2/2+\left(v_3\pm (m-\eta)\right)^2}{\rho^2} \pm 2\eta\frac{v_3^\prime}{\rho}
	+\frac{\sigma^2u_1^2}{2}+(\sigma u_3\mp\kappa)^2+\frac{g^2}{2}(f_1^2+f_2^2)	\, ,\notag \\
\Dhopm &	= \frac{v_1^2/4+ \left(\frac{v+v_3}{2}\mp m\right)^2}{\rho^2}
		+\left(\frac{\sigma u_1}{2} \right)^2+\left(\frac{\sigma}{2}(u+u_3)\pm \kappa\right)^2	
		+\frac{\beta}{4}(2f_1^2+f_2^2-1)	\notag \\
		& +\frac{f_1^2}{4}+\frac{g^2f_2^2}{2}	\, ,\notag \\
\Dhtpm &	= \frac{v_1^2/4+ \left(\frac{v-v_3}{2}\mp m\right)^2}{\rho^2}
		+\left(\frac{\sigma u_1}{2} \right)^2+\left(\frac{\sigma}{2}(u-u_3)\pm \kappa\right)^2	
		+\frac{\beta}{4}(f_1^2+2f_2^2-1)	\notag \\
		& +\frac{f_2^2}{4}+\frac{g^2f_1^2}{2}	\, ,\notag \\
\mathcal{Q} ~~&	= -\sqrt{2}\sigma u_3^\prime	\, ,\, \, \,
V_{1,2} 	= (1-\beta)\frac{f_{1,2}^2}{4}	\, ,\, \, \,
V_0   ~ 	= (1-\beta)\frac{f_1f_2}{4}	\, ,\notag \\
V_\pm ~ &	= \frac{v_1}{\rho^2}\left(\frac{v}{2}\mp m\right)
		+\sigma u_1\left(\frac{\sigma u}{2}\pm\kappa\right)+(g^{\prime2}-g^2+\beta)\frac{f_1f_2}{4}\, .
\end{align}
The vector-vector couplings are defined by 
\begin{align}\label{MATRIX_COUPL0_DEF}
\Gxy&=\left( \begin{array}{ccc} 
\dxyz	&\exyp	&\exym	\\	
 \exym	&\dxyp	&0	\\	
 \exyp	&0	&\dxym	
\end{array}\right)	\, , 
\end{align}
where $x$ and $y$ design $\mathcal{Z}$, $\mathcal{A}$, $\mathcal{W}^+$, $\mathcal{W}^-$ and 
\begin{align}\label{OPERATOR_COUPL0}
\dzae ~&	= -gg^\prime\left(\frac{v_1^2}{\rho^2}+\sigma^2u_1^2+\left(g^2-g^{\prime2}\right)f_2^2 \right)	\, ,\, \, \,
\dwwe ~= -\frac{1}{2}\left(\frac{v_1^2}{\rho^2}+\sigma^2u_1^2 \right)	\, ,\notag \\
\dzwpme &	= -g\sqrt{2}\left( \pm\eta\frac{v_1^\prime}{\rho} 
		+\frac{v_1}{\rho^2}\left(\frac{v_3}{2}\pm(m-\eta)\right)
		+\sigma u_1\left(\frac{\sigma u_3}{2}\mp\kappa\right)-\frac{g^{\prime2}}{2}f_1f_2 \right)	\, ,\notag \\
\dawpme &	= g^\prime\sqrt{2}\left( \pm\eta\frac{v_1^\prime}{\rho} 
		+\frac{v_1}{\rho^2}\left(\frac{v_3}{2}\pm(m-\eta)\right)
		+\sigma u_1\left(\frac{\sigma u_3}{2}\mp\kappa\right)+\frac{g^2}{2}f_1f_2 \right)	\, ,
\end{align}
while 
\begin{align}\label{OPERATOR_COUPL00}
\ezwpme &	= \pm g\left(\sigma u_1^\prime\pm\eta\frac{\sigma}{\rho}(v_3u_1-v_1u_3) \right)	\, ,&
\ezae &	= 0	\, ,\notag \\
\eawpme &	= \mp g^\prime\left(\sigma u_1^\prime\pm\eta\frac{\sigma}{\rho}(v_3u_1-v_1u_3) \right)	\, ,&
\ewwe &	= 0	\, .
\end{align}
Finally, the vector-scalar couplings are
\begin{align}\label{MATRIX_COUPL1_DEF}
\Gzh&=\left( \begin{array}{cccc} 
-\aoz	&\aoz	&(g^2-g^{\prime 2})\atz	&(g^{\prime 2}-g^2)\atz	\\	
 \aop	&\aom	&(g^{\prime 2}-g^2)\atp	&(g^{\prime 2}-g^2)\atm	\\	
 \aom	&\aop	&(g^{\prime 2}-g^2)\atm	&(g^{\prime 2}-g^2)\atp		
\end{array}\right)	\, ,&
\Gah&=2gg^\prime\left( \begin{array}{cccc} 
 0	&0	&-\atz	&\atz	\\	
 0	&0	&\atp	&\atm	\\	
 0	&0	&\atm	&\atp		
\end{array}\right)	\, ,
\notag \\ & & \notag \\
\Gwph&=g\sqrt{2}\left( \begin{array}{cccc} 
 0	&\atz	&-\aoz	&0	\\	
 0	&\atm	&\aop	&0	\\	
 0	&\atp	&\aom	&0	\\	
\end{array}\right)	\, ,&
\Gwmh&=g\sqrt{2}\left( \begin{array}{cccc} 
-\atz	&0	&0	&\aoz	\\	
 \atp	&0	&0	&\aom	\\	
 \atm	&0	&0	&\aop		
\end{array}\right)	\, ,
\end{align}
where
\begin{align}\label{OPERATOR_COUPL1}
\aoz &	= \frac{\sigma}{2}\left((u+u_3)f_1+u_1f_2 \right)	, &
\atz &	= \frac{\sigma}{2}\left((u-u_3)f_2+u_1f_1 \right)	, \\
\aopm &	= \frac{1}{\sqrt{2}}\left(f_1^\prime\pm\frac{1}{2\rho}\left((v+v_3)f_1+v_1f_2\right) \right)	, &
\atpm &	= \frac{1}{\sqrt{2}}\left(f_2^\prime\pm\frac{1}{2\rho}\left((v-v_3)f_2+v_1f_1\right) \right).\notag 
\end{align}

\end{document}